\documentclass[10pt]{article}
\usepackage[margin=1in]{geometry}
\usepackage[T1]{fontenc}
\usepackage{lmodern}
\usepackage{amsmath,amssymb,amsthm}
\usepackage{booktabs,array,longtable}
\usepackage{tikz}
\usetikzlibrary{arrows.meta,decorations.pathreplacing,positioning,fit,shapes.geometric,calc}
\usepackage[hidelinks,pdftitle={Pair-Merge Tokenisation Is Hard to Approximate on Binary Strings and Strongly NP-Hard on Unary Datasets},pdfauthor={Kundan Baliga, Aiden Xie, Aditya Byju}]{hyperref}
\usepackage[capitalise,noabbrev]{cleveref}
\crefname{appendix}{Appendix}{Appendices}
\crefname{subappendix}{Appendix}{Appendices}
\makeatletter\g@addto@macro{\UrlBreaks}{\do\_\do\.}\makeatother
\usepackage{thmtools}
\renewcommand\thepart{\Roman{part}}
\newcommand\mypart[1]{\par\bigskip\refstepcounter{part}\pdfbookmark[0]{Part \thepart. #1}{part.\thepart}%
  \begin{center}\rule{0.6\textwidth}{0.4pt}\\[4pt]\Large\bfseries Part~\thepart.\enspace #1\end{center}\par}
\crefname{part}{Part}{Parts}

\declaretheorem[name=Theorem]{theorem}
\declaretheorem[name=Lemma,sibling=theorem]{lemma}
\declaretheorem[name=Corollary,sibling=theorem]{corollary}
\declaretheorem[name=Proposition,sibling=theorem]{proposition}
\declaretheorem[name=Fact,sibling=theorem]{fact}
\declaretheorem[name=Conjecture,sibling=theorem]{conjecture}
\declaretheorem[name=Definition,sibling=theorem,style=definition]{definition}
\declaretheorem[name=Remark,sibling=theorem,style=definition]{remark}
\declaretheorem[name=Example,sibling=theorem,style=definition]{example}

\newcommand{\LENf}{\mathrm{LEN}_{\mathrm{full}}}
\newcommand{\LENp}{\mathrm{LEN}_{\mathrm{part}}}

\newcommand{\vc}{\mathrm{vc}}
\newcommand{\val}{\mathrm{val}}

\newcommand{\sched}{\mathrm{sched}}
\newcommand{\gl}[1]{\triangleleft_{#1}}
\newcommand{\gr}[1]{\triangleright_{#1}}
\newcommand{\Ccode}{C_{\mathrm{code}}}
\newcommand{\Cbar}{\bar C}
\newcommand{\cost}{\mathrm{cost}}

\newcommand{\str}[1]{\texttt{#1}}

\newcommand{\ourcite}[1]{\cite{#1}}
\title{Pair-Merge Tokenisation Is Hard to Approximate on Binary Strings\\ and Strongly NP-Hard on Unary Datasets}
\author{Kundan Baliga\\{\small University of Illinois Urbana-Champaign}\\{\small\texttt{kbaliga2@illinois.edu}}
\and Aiden Xie\\{\small Georgia Institute of Technology}\\{\small\texttt{axie82@gatech.edu}}
\and Aditya Byju\\{\small Georgia Institute of Technology}\\{\small\texttt{abyju6@gatech.edu}}}
\date{}

\begin{document}
\maketitle

\begin{abstract}
Tokenisers such as Byte-Pair Encoding (BPE) shorten text by repeatedly merging an adjacent pair of symbols into a new one; fewer tokens mean less computation. We ask whether an efficient algorithm can choose $k$ merges that leave a string of length $n$ almost as short as possible; doing nothing is already within a factor $n$. We prove that, unless $\mathrm P=\mathrm{NP}$, no polynomial-time algorithm comes within a factor $c\,n/\log n$, for some constant $c>0$, already for one binary string and for BPE-style merges that replace every occurrence. The key idea is that $k$ merges form a grammar with $k$ rules: a vertex-cover gap becomes a grammar-size gap, which we amplify with relabelled copies, encode in binary and realise with explicit merges. For merges that may skip occurrences, a simple algorithm achieves $O(n/\log\log n)$, and no Karp reduction can prove hardness beyond order $n/\log n$ unless $\mathrm{coNP}\subseteq\mathrm{NP/poly}$. Over a one-letter alphabet, both tokenisation problems left open by Kastreva et al.\ are strongly NP-complete on datasets with repeated strings.
\end{abstract}

\section{Introduction}\label{sec:intro}

Most language models read text through a tokeniser built by Byte-Pair Encoding (BPE) \cite{gage1994,sennrich2016bpe}.
BPE starts from characters or bytes and repeats one step: pick an adjacent pair of symbols, give it a new name, and
replace the pair by the new symbol. Fewer tokens mean less compute, so a natural question is: given a string and a
budget of $k$ merges, which merges make the string shortest, and can an efficient algorithm get close? The
\emph{compressed length} counts the symbols that remain. We compare an algorithm's compressed length with the optimum;
for a string of length $n$, doing nothing is within a factor $n$, because the optimum is at least one symbol.

We study two kinds of merge. A \emph{full merge} replaces every occurrence of its pair, as BPE does. A \emph{partial
merge} replaces a chosen set of occurrences. From here on we use only these two terms. Kozma and Voderholzer asked
whether any polynomial-time algorithm approximates the compressed length well \cite[Section~6]{kv2026arxiv}; over
alphabets that grow with the input, \cref{thm:unbounded} answers this up to a constant factor.
Kastreva, Whittington, Komm and Pimentel asked whether the binary versions of their tokenisation problems lie in APX,
that is, admit some constant-factor approximation, and left two decision problems over a one-letter alphabet open
\cite[Section~6]{kastreva2026bounded}. This paper addresses these questions for bounded alphabets, except for direct
tokenisation (\cref{sec:limits}). For the question of Kozma and Voderholzer it narrows the answer to between $n/\log n$
and $n/\log\log n$ for partial merges; for full merges the upper end is $n/(\log\log n)^{\log_32}$.

\begin{center}
\fbox{\begin{minipage}{0.94\textwidth}
\textbf{Main results.} Here $n$ is the length of the input string (or the total length of a dataset).
\begin{enumerate}
\item \emph{Binary strings: hard to approximate} (\cref{thm:main,cor:kastreva}). Unless $\mathrm P=\mathrm{NP}$, no
  polynomial-time algorithm approximates the compressed length after $k$ full or partial merges within $c_\star n/\log n$,
  for a constant $c_\star>0$, already for one binary string. So binary bottom-up tokenisation is not in APX. (For partial
  merges this follows from bounds of Nowicki \cite{nowicki2026binary}; for full merges, as in BPE, it is new.)
\item \emph{One-letter datasets: strongly NP-complete} (\cref{thm:unary,thm:unaryfull}). Over the alphabet $\{\str a\}$,
  standard bottom-up tokenisation and tokenisation by merges followed by optimal segmentation are NP-complete even when
  all string lengths are written in unary, on datasets with repeated strings. This settles both open unary problems of
  Kastreva et al.
\item \emph{How far hardness proofs can go} (\cref{part:algo}). For partial merges a simple algorithm achieves
  $O(n/\log\log n)$, and no Karp reduction (the usual form of an NP-hardness proof) proves hardness for a ratio
  $\omega(n/\log n)$ unless $\mathrm{coNP}\subseteq\mathrm{NP/poly}$, a collapse that complexity theorists consider
  unlikely. So, for partial merges, result~1 is optimal up to the
  constant among Karp reductions.
\end{enumerate}
\emph{Also in the paper:} the $\Theta(n)$ bound over alphabets that grow with the input, with its full proof
(\cref{thm:unbounded}); an $n^{1-\varepsilon}$ version of result~1 without the PCP theorem
(\cref{thm:main}(i)); for full merges, analogues of result~3 with $\log n$ and $\log\log n$ raised to the power $\log_32\approx0.63$ (\cref{thm:omsupper,cor:algo,thm:barrier}); and a closed formula for
one unary string under full merges (\cref{lem:single}).
\end{minipage}}
\end{center}

\paragraph{Terms from complexity theory.} A problem is in \emph{APX} if some polynomial-time algorithm approximates it
within a constant factor. A problem with numbers in its input is \emph{strongly} NP-hard if it stays NP-hard when all
numbers are written in unary; for one-letter strings this rules out hardness that comes only from exponentially long
strings. $\mathrm{coNP}\subseteq\mathrm{NP/poly}$ would mean that unsatisfiable formulas have short certificates
given advice that depends only on the input length; it would collapse the polynomial hierarchy.

Binary strings are strings over every larger alphabet, so the hardness holds for any alphabet with at least two
letters, including bytes. The hard instances are worst cases with a large budget, $k=\Theta(n/\log n)$: in the yes case
it compresses the whole string to one token. \Cref{sec:limits} explains why this does not say that BPE behaves badly on
natural text.

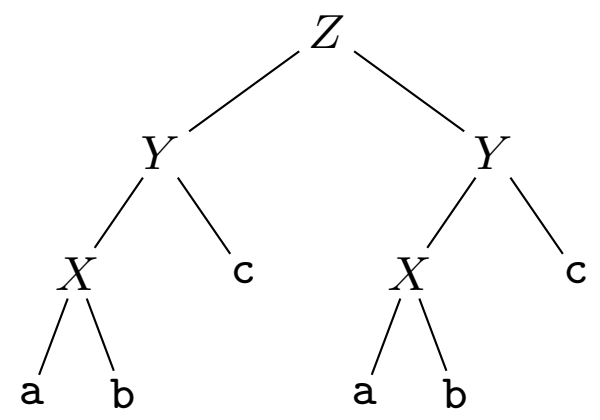
\begin{figure}[!htb]
\centering
\begin{minipage}[c]{0.5\textwidth}\small
\renewcommand{\arraystretch}{1.15}
\begin{tabular}{@{}ll@{}}
\toprule
merge & sequence\\
\midrule
start & $\str a\,\str b\,\str c\,\str a\,\str b\,\str c$\\
$(\str a,\str b)\to X$ & $X\,\str c\,X\,\str c$\\
$(X,\str c)\to Y$ & $Y\,Y$\\
$(Y,Y)\to Z$ & $Z$\\
\bottomrule
\end{tabular}
\end{minipage}%
\begin{minipage}[c]{0.4\textwidth}
\centering
\begin{tikzpicture}[level distance=8mm,level 1/.style={sibling distance=22mm},level 2/.style={sibling distance=11mm},
  level 3/.style={sibling distance=6mm},every node/.style={font=\small,inner sep=1.5pt}]
\node {$Z$}
  child {node {$Y$} child {node {$X$} child {node {$\str a$}} child {node {$\str b$}}} child {node {$\str c$}}}
  child {node {$Y$} child {node {$X$} child {node {$\str a$}} child {node {$\str b$}}} child {node {$\str c$}}};
\end{tikzpicture}
\end{minipage}
\caption{The two bridges on $s=\str{abcabc}$. Left: three merges reduce $s$ to one symbol. Right: read as rules
$X\to\str{ab}$, $Y\to X\str c$, $Z\to YY$, the same merges form a grammar, whose parse tree is shown
(\cref{lem:merge2slp}). Conversely, processing the rules of any grammar children first gives partial merges
(\cref{lem:realise}). A rule at most doubles the length of what it derives, so $g(s)=3$, and by \cref{lem:merge2slp}
two merges leave at least $3-2+1=2$ symbols.}\label{fig:bridge}
\end{figure}

\begin{figure}[!htb]
\centering
\begin{tikzpicture}[>=Stealth, box/.style={draw,rounded corners=2pt,align=center,inner sep=4pt,font=\small,minimum height=10mm,text width=2.75cm},
  lab/.style={font=\footnotesize,align=center}]
\node[box] (G) {graph $H$\\ vertex-cover gap $\Delta$};
\node[box,right=1.15cm of G] (S) {gadget string $s_H$\\ grammar gap $\Delta$};
\node[box,right=1.15cm of S] (W) {$t$ affine copies\\ gap $t\Delta$};
\node[box,right=1.15cm of W] (F) {binary string $F$\\ gap $\ge t\Delta-\Cbar$};
\draw[->] (G) -- (S); \draw[->] (S) -- node[lab,above]{copies} (W); \draw[->] (W) -- node[lab,above]{code} (F);
\end{tikzpicture}
\caption{The plan of the binary hardness proof. Merges are grammar rules, so the vertex-cover gap becomes an additive
gap in grammar size; the copies multiply it by $t$. The binary code costs $\Cbar=O(p\,|V|\log|V|)$ rules for cubic $H$
and field size $p$, while the gap grows like $t=p^2$, so for large $p$ the code cost is negligible. In the yes case an explicit schedule of full merges reduces $F$ to one token; in the no case many tokens remain.}\label{fig:keyidea}
\end{figure}
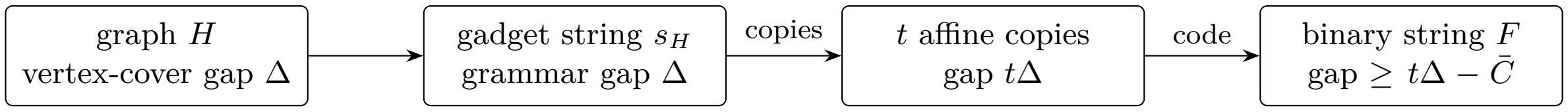

\paragraph{Key idea: merges are small grammars.} A sequence of $k$ merges is a grammar with $k$ rules $X\to YZ$, much
like a BPE merge table. So if every grammar for a string needs $k+\Delta$ rules, then $k$ merges leave at least
$\Delta+1$ symbols; and if a grammar with $k$ rules exists and the merges can follow it, they leave one symbol
(\cref{fig:bridge}). Because the yes case leaves exactly one symbol, the approximation ratio is simply the number of
symbols left in the no case, so an additive gap in grammar size becomes a multiplicative gap in compressed length. We start from a string over a large alphabet whose smallest grammar size encodes the vertex-cover
number of a graph. \Cref{sec:warmup} carries out this whole argument on a graph with one edge, small enough to check
by hand. Writing that string over a fixed alphabet naively destroys the gap, so we follow Nowicki
\cite{nowicki2026binary} (\cref{fig:keyidea}): take many copies of the string, relabel each copy by a different affine
map so that no two copies share an adjacent pair, and write every symbol in a binary code. The encoding is paid once,
while the vertex-cover gap is multiplied by the number of copies. With $t$ copies of a gadget string of length
$\Theta(|V|)$ and a gap $\Delta=\Theta(|V|)$, the binary string has $n=\Theta(t\,|V|\log n)$ bits, the codeword length
being $\Theta(\log n)$, and $\Omega(t|V|)=\Omega(n/\log n)$ symbols remain in the no case. Partial merges can follow any grammar; full merges
cannot, because they also replace unwanted occurrences. Our new step is an explicit schedule of full merges for the
binary string (\cref{sec:binary-upper}).

\paragraph{Key idea of the unary part.} With one letter, a string is just a number and a token is a number too. We
write every number in the construction as a set of bits. A vertex is ``chosen'' by merging its binary code with a middle
bit, and one last merge of the two middle bits brings every other edge string to $3$ tokens. So an edge string reaches
the minimum of $2$ tokens only if both endpoints were chosen: an AND gadget (\cref{fig:biclique}),
which encodes balanced biclique. Many copies of auxiliary strings force every optimal solution to build the codes
first, at a known cost. The merges-then-segmentation result uses a different, number-theoretic construction from vertex
cover (\cref{sec:unary}).

\paragraph{What is new.} For partial merges the binary hardness follows from Nowicki's bounds on binary grammar size,
which he states only for exact hardness. Our contribution in the binary part is the full-merge schedule and hence the
answer to Kastreva et al.'s APX question for bottom-up tokenisation (the direct half stays open). The Karp barrier and
the full-merge algorithm of \cref{part:algo} are new; the partial-merge algorithm is elementary. The unary part is new:
the structure of full merges on one-letter strings, the AND gadget and the scaffold constructions. \Cref{sec:related}
compares with prior work in detail.

\paragraph{Roadmap.} \Cref{sec:prelim} gives definitions and the two bridges between merges and grammars.
\Cref{part:hardness} contains the hardness results: the hand-checkable warm-up over a large alphabet (\cref{sec:warmup}),
the binary hardness (\cref{sec:binary}) and the unary results (\cref{sec:unary}). \Cref{part:algo} (\cref{sec:gap})
gives the algorithms and the barrier. \Cref{part:discussion} covers related work and connections (\cref{sec:related})
and scope, practice and open problems (\cref{sec:limits}). Case checks, routine lemmas and all computer checks are in
the appendices.

\section{Definitions and two bridges between merges and grammars}\label{sec:prelim}

\subsection{Merges}

A string $s$ of length $n=|s|$ is a sequence of symbols. A \emph{merge} of a pair $(a,b)$ of symbols introduces a
fresh symbol $X$ and replaces occurrences of the adjacent pair $ab$ in the current sequence by $X$. We follow Kozma
and Voderholzer \cite{kv2026arxiv}, and call the symbols of the current sequence \emph{tokens}.
\begin{itemize}
\item A \emph{full merge} replaces all occurrences, scanning left to right, without overlaps. For example, the full
  merge of $(\str a,\str a)$ turns $\str{aaa}$ into $X\str a$. This is what BPE does. The resulting problem is OMS (optimal
  merge sequence), the problem of Zouhar et al.\ \cite{zouhar2023formal} in the notation of \cite{kv2026arxiv}.
\item A \emph{partial merge} replaces any set of pairwise non-overlapping occurrences, possibly none. The resulting
  problem is OPE (optimal pair encoding).
\end{itemize}
A merge whose pair does not occur changes nothing; we call it an \emph{identity merge}. So ``exactly $k$ merges'' and
``at most $k$ merges'' give the same optimum; in the string-identified model of Kastreva et al.\ below, a merge of two
strings that never occur adjacently plays the same role. The budget $k$ is part of the input. Let $\LENf(s,k)$ and $\LENp(s,k)$ be the minimum length of the sequence left after $k$ full,
respectively partial, merges. A full merge is a special partial merge, so
\begin{equation}\label{eq:pf}
1\le \LENp(s,k)\le \LENf(s,k)\le n .
\end{equation}
The \emph{utility} of a merge sequence is $n$ minus the final length, the number of symbols saved.

An algorithm \emph{approximates the compressed length within a factor $\rho\ge1$} if on every input $(s,k)$ it
outputs $k$ merges whose final length is at most $\rho$ times the optimum. By \eqref{eq:pf}, doing nothing is an
$n$-approximation.

\paragraph{Datasets and tokenisers.} Kastreva et al.\ \cite{kastreva2026bounded} work with a dataset $\mathcal
D$ (a multiset of strings) and the objective $\sum_{u\in\mathcal D}|\mathrm{tok}(u)|$, the total number of tokens; for
a dataset, $n$ denotes the total length $\sum_{u\in\mathcal D}|u|$. Their budget, written $K$ there, is our $k$. Their
\emph{bottom-up} tokeniser applies $k$ merges to every string, each replacing all occurrences left to right, as in full
merges; their tokens are identified by the strings they spell rather than by fresh names (\cref{sec:binary-final}). Their
OPE tokeniser, which we call \emph{OPE-seg}, takes the tokens produced by $k$ merges and segments every string optimally
with that vocabulary. Their \emph{direct} tokeniser chooses any vocabulary of $k$ extra tokens and segments optimally.
\Cref{tab:names} maps these names.

\begin{table}[!htb]
\centering\footnotesize
\setlength{\tabcolsep}{3.5pt}
\begin{tabular}{@{}lllll@{}}
\toprule
name here & Kastreva et al. & merges act on & output & practical analogue\\
\midrule
full merges (OMS) & bottom-up & all occurrences, in order & token string & BPE\\
partial merges (OPE) & -- & chosen occurrences & token string & choose where merges apply\\
OPE-seg & ``OPE'' & (define a vocabulary) & optimal segmentation & BPE vocab., best segmentation\\
direct (not covered) & direct & (any $k$ tokens) & optimal segmentation & Unigram/WordPiece-like\\
\bottomrule
\end{tabular}
\caption{Problem names. OPE is Kozma and Voderholzer's partial-merge problem; OPE-seg is the tokeniser of Kastreva et
al.\ that segments optimally with the vocabulary of $k$ merges.}\label{tab:names}
\end{table}

\subsection{Straight-line programs}

A \emph{binary straight-line program} (SLP) is an acyclic set of rules $X\to YZ$, where $Y$ and $Z$ are letters or
nonterminals. Each nonterminal $X$ derives a string $\val(X)$. For $|s|\ge2$, the \emph{grammar size} $g(s)$ is the
minimum number of rules of an SLP that has a nonterminal deriving $s$. For a single letter, $g=0$. For a finite set
$\mathrm{Words}$ of words of length at least $2$, $g(\mathrm{Words})$ is the minimum number of rules of an SLP in which
every word of $\mathrm{Words}$ is derived by some nonterminal. Without loss of generality distinct nonterminals have
distinct values (identifying two nonterminals with equal values never adds rules), so we may name a nonterminal by
its value. The \emph{parse tree} of an SLP for $s$ unfolds the rules from the nonterminal deriving $s$; each node
covers an interval of $s$, its \emph{yield}.

Over a fixed alphabet every string of length $n$ has $g=O(n/\log n)$ \cite{kieffer2000grammar}. We use this only to
explain why our binary ratio has a $\log n$ in it.

\begin{table}[!tp]
\centering\small
\begin{tabular}{@{}p{0.2\textwidth}p{0.74\textwidth}@{}}
\toprule
symbol & meaning\\
\midrule
$n$, $k$ & length of the input string (or total length of a dataset); merge budget\\
$\mathcal D$, $\rho$ & dataset (multiset of strings); approximation ratio\\
$\LENf,\LENp$ & minimum length after $k$ full / partial merges\\
$g(s)$, $g(\mathrm{Words})$ & smallest number of binary rules deriving $s$ / all words of $\mathrm{Words}$\\
$\val(X)$ & string derived by a nonterminal or token $X$\\
$H=(V,E)$, $\vc(H)$, $\kappa$ & graph; its vertex-cover number; threshold of the source instance\\
$S$ & a (minimal) vertex cover; $|S|$ its size\\
$\varepsilon_0$, $c_0$, $c_\star$ & gap constant of \cref{fact:cubic}; constants of \cref{thm:unbounded,thm:main}, proportional to $\varepsilon_0$\\
$\lambda$ & number of strings in a dataset (\cref{rem:lambda})\\
\midrule
$\#$, $x_v$, $\$_j$ & gadget letters: hash, vertex letter of $v\in V$, separator number $j$\\
(ALT) & adjacent letters differ, and the first letter differs from the last\\
$A_v,B_v,C_v,E_{uv}$ & tokens with values $\#x_v$, $x_v\#$, $\#x_v\#$, $\#x_u\#x_v\#$\\
$s_H$, $N$, $R(c)$ & the gadget string of $H$; its number of words, $N=2|V|+|E|$; $R(c)=6|V|+3|E|-1+c$\\
$\sched(S)$ & the full-merge schedule for $s_H$ built from a minimal cover $S$ (\cref{lem:sched})\\
\midrule
$w$, $q$ & source word for the binary construction ($w=s_H$); its number of distinct letters\\
$p$, $t$ & prime field size; number of copies, $t\le p^2$\\
$\pi_{a,b}$, $\gl a$, $\gr b$ & affine relabelling $x\mapsto(x,ax+b)$; left and right guard letters\\
$W$, $W_{a,b}$, $q_W$ & amplified string $W=W_{(1)}\cdots W_{(t)}$; one copy (block); number of distinct letters of $W$\\
$T=T_1\cdots T_r$ & a factorisation of a string into $r$ factors (\cref{sec:binary-lower})\\
$\eta,\bar\eta(i)$ & bit code $\eta(0)=\str{10}$, $\eta(1)=\str{11}$; the $2d$-bit identifier of letter $i$\\
$d$, $L$, $\chi_i$, $h$, $F$ & bits per identifier; codeword length $4d+3$; codeword of letter $i$; the coding morphism; $F=h(W)$\\
anchor & bit at offset $2d+2$ of a codeword (first bit of the second identifier copy)\\
$\Omega,\beta_0,\beta_1,\Gamma_i$ & tokens for $\str{00}$, $\str{10}$, $\str{11}$, and for the whole codeword $\chi_i$ (the token $\Omega$ is unrelated to the asymptotic $\Omega(\cdot)$)\\
$P_\varphi$, $\tau_i$, $\theta$ & Stage-A token for a codeword prefix $\varphi$; token form of $\chi_i$ after three merges; last token of a prefix $\varphi=\varphi'\theta$\\
$\Psi_{a,b},U_{a,b},U_{\le j}$ & Stage-B tokens: one copy; one copy with its guards; the first $j$ copies\\
$\Ccode$, $\Cbar$ & number of merges in Stage~A; the upper bound $3+(2d+1)q_W$ used in the reduction\\
$\Delta$ & additive vertex-cover gap $\vc(H)-\kappa$\\
\bottomrule
\end{tabular}
\caption{Notation for \cref{sec:prelim,sec:warmup,sec:binary}.}\label{tab:notation}
\end{table}

\subsection{Two bridges}

The first bridge says that merges are grammars. It turns a lower bound on grammar size into a lower bound on
compressed length.

\begin{lemma}[merges give grammars]\label{lem:merge2slp}
If $k$ partial merges turn $s$ into a sequence of length $\ell$, then $g(s)\le k+\ell-1$. Consequently, for every
$k$,
\[
\LENf(s,k)\ \ge\ \LENp(s,k)\ \ge\ g(s)-k+1 .
\]
\end{lemma}

\begin{proof}
Each merge of $(a,b)$ into $X$ gives a rule $X\to ab$. Every symbol of the final sequence $u_1\cdots u_\ell$ derives
the part of $s$ that it covers, so $\val(u_1)\cdots\val(u_\ell)=s$. Add $\ell-1$ rules that join $u_1,\dots,u_\ell$
from left to right (none if $\ell=1$). Merges with no occurrence only add unused rules. A merge of a pair $(a,a)$ is
no different. The second statement follows with \eqref{eq:pf}.
\end{proof}

The bound also holds when merges are identified by the strings they produce, as in Kastreva et al.: each merge
still contributes at most one rule.

The second bridge is the converse for partial merges. It is the reason partial merges are easy to handle and full
merges are not.

\begin{lemma}[partial merges realise any grammar]\label{lem:realise}
For every string $s$ with $|s|\ge2$, $\LENp(s,g(s))=1$.
\end{lemma}

\begin{proof}
Take a smallest SLP for $s$ with pairwise distinct values, and its parse tree. No node labelled $X$ is a proper
descendant of another node labelled $X$, since the SLP is acyclic; so distinct nodes labelled $X$ have disjoint yields.
Process the rules children first.
We keep the invariant that the current sequence is the sequence of maximal already-processed nodes of the tree.
When the rule $X\to YZ$ is processed, every node labelled $X$ has two children that are adjacent maximal processed
nodes, labelled $Y$ and $Z$. We partially merge exactly those occurrences of the pair $YZ$; they do not overlap, by the
disjointness above, and there is at least one. After $g(s)$ merges only the
root is left.
\end{proof}

Together, the two lemmas say that, for $|s|\ge2$, $\LENp(s,k)=1$ if and only if $g(s)\le k$. So, for partial merges, an
\emph{additive} gap in grammar size becomes a \emph{multiplicative} gap in compressed length: if $g(s)\le k$ in the
yes case and $g(s)\ge k+\Delta$ in the no case, then the optimum is $1$ versus at least $\Delta+1$. All our hardness
results follow this template. For full merges, \cref{lem:realise} is not available, and we must build the merges by
hand.

\subsection{Vertex cover}

For a graph $H=(V,E)$, a \emph{vertex cover} is a set $S\subseteq V$ that contains an endpoint of every edge. It is
\emph{minimal} if no proper subset is a vertex cover. A minimum vertex cover is minimal. Write $\vc(H)$ for the
minimum size. Deciding $\vc(H)\le\kappa$ is NP-complete \cite{karp1972}, also for cubic graphs (this follows from
\cref{fact:cubic}; see also \cite{gjs1976}). We also use the following gap version. In words: it is NP-hard to tell graphs
with a cover of size $\kappa$ from graphs in which every cover is a constant factor larger.

\begin{fact}[APX-hardness of cubic vertex cover \cite{alimonti2000cubic,berman1999tighter,chlebik2006bounded}]\label{fact:cubic}
There are a constant $\varepsilon_0>0$ and a polynomial-time reduction from SAT to pairs $(H,\kappa)$, with $H$ a
simple cubic graph, such that satisfiable instances give $\vc(H)\le\kappa$ and unsatisfiable instances give
$\vc(H)\ge(1+\varepsilon_0)\kappa$.
\end{fact}

These results come from gap-preserving reductions from the PCP theorem \cite{arora1998pcp}; we do not quote a
numerical value of $\varepsilon_0$, and constants derived from it below are not explicit either. In a cubic graph
$|E|=3|V|/2$ and every vertex covers $3$ edges, so $\vc(H)\ge|V|/2$. So when $\kappa<|V|/2$ the instance is a
no-instance, and our reductions map it to a fixed no-instance; below we assume $\kappa\ge|V|/2$.

\subsection{Notation}

\Cref{tab:notation} lists the symbols of \cref{sec:prelim,sec:warmup,sec:binary}; within these sections each has one
meaning. \Cref{sec:gap,sec:unary} are self-contained and declare their own local symbols at their start, some of
which reuse letters from the table with a different, section-local meaning.

\mypart{Hardness}\label{part:hardness}
\section{Warm-up: the whole argument over a large alphabet}\label{sec:warmup}

This section proves the simplest version of our result: over an alphabet that grows with the input, unless
$\mathrm P=\mathrm{NP}$, the compressed length cannot be approximated within $c_0n$ for a constant $c_0>0$. It uses only the two bridges of \cref{sec:prelim} and, for
the final constant, \cref{fact:cubic}. Every step can be checked by hand on the smallest graph, a single edge, which we
work through as we go (\cref{ex:k2,fig:witness,fig:k2}). The argument first appeared in an earlier version of our preprint on
byte-pair encoding \ourcite{baliga2026bpevonetwo}; this paper is now its home. We give it in full because the binary
construction uses its gadget, its grammar lemma and its merge schedule as black boxes.

\subsection{The gadget}

Let $H=(V,E)$ be a graph without isolated vertices, with $V=\{1,\dots,|V|\}$ and each edge written $(u,v)$ with
$u<v$. The alphabet has a letter $\#$, a letter $x_v$ for every vertex $v$, and separators $\$_1,\$_2,\dots$. The
\emph{words} are
\[
\#x_v,\qquad x_v\#\qquad(v\in V),\qquad\qquad \#x_u\#x_v\#\qquad((u,v)\in E).
\]
There are $N=2|V|+|E|$ words. The gadget string $s_H$ lists the words in a fixed order, each followed by its own
separator: $s_H=w_1\$_1w_2\$_2\cdots w_N\$_N$. Its length is $6|V|+6|E|$ and it has $q=1+|V|+N$ distinct letters.
Two properties matter later. In $s_H$, adjacent symbols are always different, and the first symbol differs
from the last, which is the separator $\$_N$. We call this \textbf{property (ALT)}.

\begin{example}[the single edge $K_2$]\label{ex:k2}
For $V=\{1,2\}$ and $E=\{(1,2)\}$ we have $N=5$ words and
\[
s_{K_2}=\#x_1\,\$_1\ \#x_2\,\$_2\ x_1\#\,\$_3\ x_2\#\,\$_4\ \#x_1\#x_2\#\,\$_5,
\]
a string of length $18$ over $8$ letters. See \cref{fig:k2}.
\end{example}

\subsection{Grammar size encodes the vertex-cover number}

\paragraph{Intuition.} Unique separators cannot be shared, so a grammar for $s_H$ must build each word
separately and then join the $2N$ pieces. The words $\#x_v$ and $x_v\#$ cost one rule each. An edge word $\#x_u\#x_v\#$
costs one more rule, provided some nonterminal of length three or four is already there to build it from. The cheapest
such helpers are $\#x_v\#$ for the vertices of a vertex cover: one helper per cover vertex serves all edges at that
vertex.

\begin{lemma}[separator lemma]\label{lem:sep}
Let $w_1,\dots,w_N$ be words of length at least $2$ over an alphabet $\Sigma$, let $\$_1,\dots,\$_N$ be distinct
letters not in $\Sigma$, and let $\mathrm{Words}=\{w_1,\dots,w_N\}$. Then $g(w_1\$_1\cdots w_N\$_N)=2N-1+g(\mathrm{Words})$.
\end{lemma}

\begin{proof}
($\le$) Build every $w_j$ with $g(\mathrm{Words})$ rules, then join the $2N$ symbols $w_1,\$_1,\dots,w_N,\$_N$ with
$2N-1$ rules.

($\ge$) Fix a smallest SLP for $s=w_1\$_1\cdots w_N\$_N$ and its parse tree. Call a node \emph{cross} if its yield
contains a separator. (i) A cross nonterminal labels exactly one node: two nodes with the same label are either
nested, which acyclicity forbids, or have disjoint yields, which would both contain the same separator, which occurs
only once. (ii) Ancestors of cross nodes are cross, so the cross nodes form a subtree containing the root. Every child
of a cross node is cross or a \emph{piece}: a separator leaf or a non-cross node. A non-cross yield is contiguous and
contains no separator, so it lies inside one word. The pieces partition $s$ into the $N$ separators and, for each
$j$, a factorisation of $w_j$ into $f_j\ge1$ pieces. The cross subtree is a full binary tree with $N+\sum_jf_j$
leaves, so it has $N+\sum_jf_j-1$ internal nodes, which by (i) are distinct rules. The non-cross rules form an SLP
over $\Sigma$ in which each $w_j$ factorises into $f_j$ symbols; adding $f_j-1$ rules per word makes every word a
nonterminal. So the number of non-cross rules is at least $g(\mathrm{Words})-\sum_j(f_j-1)$. In total
$g(s)\ge N+\sum_jf_j-1+g(\mathrm{Words})-\sum_j(f_j-1)=2N-1+g(\mathrm{Words})$.
\end{proof}

\begin{lemma}[gadget lemma]\label{lem:gadget}
For the words $\mathrm{Words}$ of $s_H$, $g(\mathrm{Words})=2|V|+|E|+\vc(H)$.
\end{lemma}

\begin{proof}
($\le$) Let $S$ be a vertex cover. Use the rules $A_v\to\#x_v$ and $B_v\to x_v\#$ for all $v$; $C_v\to A_v\#$ for
$v\in S$, so $\val(C_v)=\#x_v\#$; and for each edge $E_{uv}\to C_uB_v$ if $u\in S$, otherwise $E_{uv}\to A_uC_v$
(then $v\in S$). That is $2|V|+|S|+|E|$ rules.

($\ge$) Any feasible SLP has nonterminals with the $2|V|+|E|$ distinct values $\#x_v$, $x_v\#$ and $\#x_u\#x_v\#$, of
lengths $2$ and $5$. The rule for $\#x_u\#x_v\#$ splits it at one of four points. Each split needs a nonterminal with
value $x_u\#x_v\#$, $\#x_v\#$, $\#x_u\#$ or $\#x_u\#x_v$, of length $3$ or $4$, so not one of the forced values. Call it
the \emph{witness} of the edge. Map witness values to vertices: $\#x_v\#\mapsto v$, $x_u\#x_v\#\mapsto u$ and
$\#x_u\#x_v\mapsto u$. Every edge's witness maps to an endpoint of that edge. So the images of all witnesses form a
vertex cover, and there are at least $\vc(H)$ distinct witnesses.
\end{proof}

\begin{figure}[!htb]
\centering\small
\renewcommand{\arraystretch}{1.15}
\begin{tabular}{@{}lll@{}}
\toprule
last rule splits $\#x_u\#x_v\#$ as & witness (length $3$ or $4$) & vertex it names\\
\midrule
$\#\ \mid\ x_u\#x_v\#$ & $x_u\#x_v\#$ & $u$\\
$\#x_u\ \mid\ \#x_v\#$ & $\#x_v\#$ & $v$\\
$\#x_u\#\ \mid\ x_v\#$ & $\#x_u\#$ & $u$\\
$\#x_u\#x_v\ \mid\ \#$ & $\#x_u\#x_v$ & $u$\\
\bottomrule
\end{tabular}
\caption{The witness argument of \cref{lem:gadget}. Whatever rule builds an edge word, one of its two parts has length
$3$ or $4$ and is not one of the forced words. This witness names an endpoint of the edge, so the witnesses name a
vertex cover. For $K_2$ (one edge, $u=1$, $v=2$) one witness is needed, and the cover $\{1\}$ provides it as
$\#x_1\#$.}\label{fig:witness}
\end{figure}

\begin{corollary}\label{cor:gsH}
$g(s_H)=R(\vc(H))$, where $R(c):=6|V|+3|E|-1+c$.
\end{corollary}

\begin{proof}
\Cref{lem:sep,lem:gadget}: $2N-1+2|V|+|E|+\vc(H)=6|V|+3|E|-1+\vc(H)$.
\end{proof}

For $K_2$ this gives $g(s_{K_2})=R(1)=12+3-1+1=15$.

\subsection{Full merges suffice in the yes case}

\paragraph{Intuition.} \Cref{lem:realise} turns the grammar of \cref{lem:gadget} into partial merges. For full
merges we must check that each merge does not also hit occurrences we do not want. The danger is the pair $(\#,x_v)$,
which occurs both in the vertex word $\#x_v$ and inside edge words. The schedule below merges cover vertices first,
so that inside an edge word the cover endpoint grabs its neighbouring $\#$ before anything else can.

\begin{lemma}[full-merge schedule]\label{lem:sched}
Let $H$ have no isolated vertices and let $S$ be a minimal vertex cover. The following sequence $\sched(S)$ of exactly
$R(|S|)$ full merges turns $s_H$ into a single symbol. Every merge has at least one occurrence.
\begin{enumerate}
\item $(\#,x_v)\to A_v$ for $v\in S$;
\item $(A_v,\#)\to C_v$ for $v\in S$;
\item $(x_v,\#)\to B_v$ for all $v\in V$;
\item $(\#,x_v)\to A_v$ for $v\notin S$;
\item for each edge $(u,v)$: $(A_u,C_v)\to E_{uv}$ if $v\in S$, otherwise $(C_u,B_v)\to E_{uv}$;
\item the spine: $2N-1$ times, merge the first two symbols of the current sequence.
\end{enumerate}
\end{lemma}

\begin{proof}
Separators occur once, so only pairs inside a word can repeat. Trace an edge
word $\#x_u\#x_v\#$:
\begin{itemize}
\item $u,v\in S$: step 1 gives $A_uA_v\#$; step 2 gives $A_uC_v$ (there is no pair $A_u\#$); steps 3 and 4 do
  nothing; step 5 merges $(A_u,C_v)$.
\item $u\in S$, $v\notin S$: step 1 gives $A_u\#x_v\#$; step 2 gives $C_ux_v\#$; step 3 gives $C_uB_v$; step 5.
\item $u\notin S$, $v\in S$: step 1 gives $\#x_uA_v\#$; step 2 gives $\#x_uC_v$; step 3 does nothing, because $x_u$
  is followed by $C_v$; step 4 gives $A_uC_v$; step 5.
\end{itemize}
The word $\#x_v$ becomes $A_v$ at step 1 or step 4 and is untouched otherwise, since it is followed by a separator.
The word $x_v\#$ is untouched by steps 1 and 2, since its $\#$ is followed by a separator, and becomes $B_v$ at step
3. Each pair in step 5 occurs exactly once, because it determines its edge. After step 5 every word $w_j$ has become a single token, which we call
$E'_j$ (one of $A_v$, $B_v$, $E_{uv}$), so the sequence is $E'_1\$_1\cdots E'_N\$_N$ with $2N$ distinct symbols, so every spine merge has exactly one occurrence.

Every merge in steps 1--5 has an occurrence. For steps 1, 3 and 4 this is the vertex words. For step 2 and $v\in S$:
by minimality $v$ has a neighbour $u\notin S$, and the trace above shows the pair $A_v\#$ in that edge word. For step
5 it is the trace. The count is $|S|+|S|+|V|+(|V|-|S|)+|E|+2N-1=R(|S|)$.
\end{proof}

Minimality of $S$ is used only to guarantee that no merge of $\sched(S)$ is an identity merge. For a non-minimal cover
the same sequence still reduces $s_H$ to one symbol, with some identity merges; this is harmless for the bounds.

\begin{figure}[!htb]
\centering
\small
\renewcommand{\arraystretch}{1.15}
\begin{tabular}{@{}l@{\quad}l@{}}
\toprule
after & sequence (cover $S=\{1\}$)\\
\midrule
start & $\#\,x_1\;\$_1\quad \#\,x_2\;\$_2\quad x_1\,\#\;\$_3\quad x_2\,\#\;\$_4\quad \#\,x_1\,\#\,x_2\,\#\;\$_5$\\
1: $(\#,x_1)\to A_1$ & $\mathbf{A_1}\;\$_1\quad \#\,x_2\;\$_2\quad x_1\,\#\;\$_3\quad x_2\,\#\;\$_4\quad \mathbf{A_1}\,\#\,x_2\,\#\;\$_5$\\
2: $(A_1,\#)\to C_1$ & $A_1\;\$_1\quad \#\,x_2\;\$_2\quad x_1\,\#\;\$_3\quad x_2\,\#\;\$_4\quad \mathbf{C_1}\,x_2\,\#\;\$_5$\\
3: $(x_1,\#)\to B_1$, $(x_2,\#)\to B_2$ & $A_1\;\$_1\quad \#\,x_2\;\$_2\quad \mathbf{B_1}\;\$_3\quad \mathbf{B_2}\;\$_4\quad C_1\,\mathbf{B_2}\;\$_5$\\
4: $(\#,x_2)\to A_2$ & $A_1\;\$_1\quad \mathbf{A_2}\;\$_2\quad B_1\;\$_3\quad B_2\;\$_4\quad C_1\,B_2\;\$_5$\\
5: $(C_1,B_2)\to E_{12}$ & $A_1\;\$_1\quad A_2\;\$_2\quad B_1\;\$_3\quad B_2\;\$_4\quad \mathbf{E_{12}}\;\$_5$\\
6: spine, $9$ merges & one symbol\\
\bottomrule
\end{tabular}
\caption{The schedule $\sched(\{1\})$ of \cref{lem:sched} on $s_{K_2}$. Bold symbols were created by the step on
that row. Note step~1: the full merge of $(\#,x_1)$ also fires inside the edge word, as intended.
Steps 1--5 use $1+1+2+1+1=6$ merges and the spine uses $9$, so $15=R(1)=g(s_{K_2})$ merges reduce the string to one
symbol. With $14$ merges, \cref{lem:merge2slp} and \cref{cor:gsH} force at least $15-14+1=2$ symbols.}\label{fig:k2}
\end{figure}

\subsection{The unbounded-alphabet theorem}

\begin{theorem}[{earlier in \ourcite{baliga2026bpevonetwo}}]\label{thm:unbounded}
There is a constant $c_0>0$ (proportional to $\varepsilon_0$) such that, unless $\mathrm P=\mathrm{NP}$, no
polynomial-time algorithm approximates $\LENp(s,k)$ or $\LENf(s,k)$ within a factor $c_0n$. Deciding whether $\LENf(s,k)=1$, and deciding whether
$\LENp(s,k)=1$, are NP-complete.
\end{theorem}

\begin{proof}
Start from \cref{fact:cubic} and map $(H,\kappa)$ with $\kappa\ge|V|/2$ to $(s_H,k)$ with $k=R(\kappa)$ (inputs with
$\kappa<|V|/2$ go to a fixed no-instance).
\begin{itemize}
\item Yes ($\vc(H)\le\kappa$): a minimum cover is minimal and a cubic graph has no isolated vertices, so
  \cref{lem:sched} reaches length $1$ with $R(\vc(H))\le R(\kappa)$ merges; pad with merges of absent pairs. So
  $\LENf=\LENp=1$.
\item No ($\vc(H)\ge(1+\varepsilon_0)\kappa$): by \cref{cor:gsH}, $g(s_H)\ge R(\kappa)+\varepsilon_0\kappa$, so by
  \cref{lem:merge2slp} $\LENf\ge\LENp\ge\varepsilon_0\kappa+1$.
\end{itemize}
Here $n=15|V|$ and $\kappa\ge|V|/2$, so $\varepsilon_0\kappa+1>\varepsilon_0n/30$. An algorithm with ratio below
$c_0n$, $c_0=\varepsilon_0/30$, would separate the two cases. For NP-membership of the decision problems: a merge with an
occurrence shortens the sequence, so at most $n-1$ merges matter and they form a certificate. Hardness is the same map
applied to the decision version of vertex cover on cubic graphs, which is NP-complete (\cref{sec:prelim}).
\end{proof}

\paragraph{What can be checked by hand.} On $K_2$ every claim of this section is a finite check: $s_{K_2}$ has
length $18$; \cref{lem:sep,lem:gadget} give $g(s_{K_2})=9+5+1=15$ (concretely: the $10$ pieces, five words and five
separators, need $9$ joining rules, and the five words need $6$ rules, one of them for the witness, which has length $3$ or $4$ and
names vertex $1$ or $2$, \cref{fig:witness}); \cref{fig:k2} shows that $15$ full merges reach one symbol, each with an occurrence; and
\cref{lem:merge2slp} shows that $14$ merges leave at least $2$ symbols. The same steps, with the cover number in place of
$1$, prove \cref{thm:unbounded}.

The hard instances have optimum $1$ and a budget $k=R(\kappa)$ that is a constant fraction of $n$. The alphabet has
$|V|$ vertex letters and $N$ separators, so it grows with $n$. Over a fixed alphabet this route needs a new idea.
If the unique separators are replaced by short strings over a fixed alphabet, every separator design we tried lets the
trailing $\#$ of each edge word be glued into a shared separator rule at constant total cost. That turns each edge word
into $A_uA_v$, which needs no witness, and the vertex-cover gap disappears. We know no proof that every design fails.
The next section gets around this by \emph{amplifying} the gap before encoding.

\section{The binary alphabet}\label{sec:binary}

\begin{theorem}[main theorem]\label{thm:main}
Unless $\mathrm P=\mathrm{NP}$, the following hold for strings over $\{0,1\}$, where $n$ is the string length.
\begin{enumerate}\renewcommand{\labelenumi}{(\roman{enumi})}
\item For every $\varepsilon>0$, no polynomial-time algorithm approximates $\LENf(s,k)$ or $\LENp(s,k)$ within a
  factor $n^{1-\varepsilon}$.
\item There is a constant $c_\star>0$ such that no polynomial-time algorithm approximates $\LENf(s,k)$ or $\LENp(s,k)$
  within a factor $c_\star n/\log n$. The constant is proportional to the (non-explicit) $\varepsilon_0$ of
  \cref{fact:cubic}.
\end{enumerate}
Part (ii) implies part (i) for large $n$. Part (i) is stated separately only because its proof uses just the
NP-hardness of vertex cover, not the PCP theorem. Both parts use the same construction with different parameters, and
in the yes case $\LENf=\LENp=1$.
\end{theorem}

\subsection{Overview}

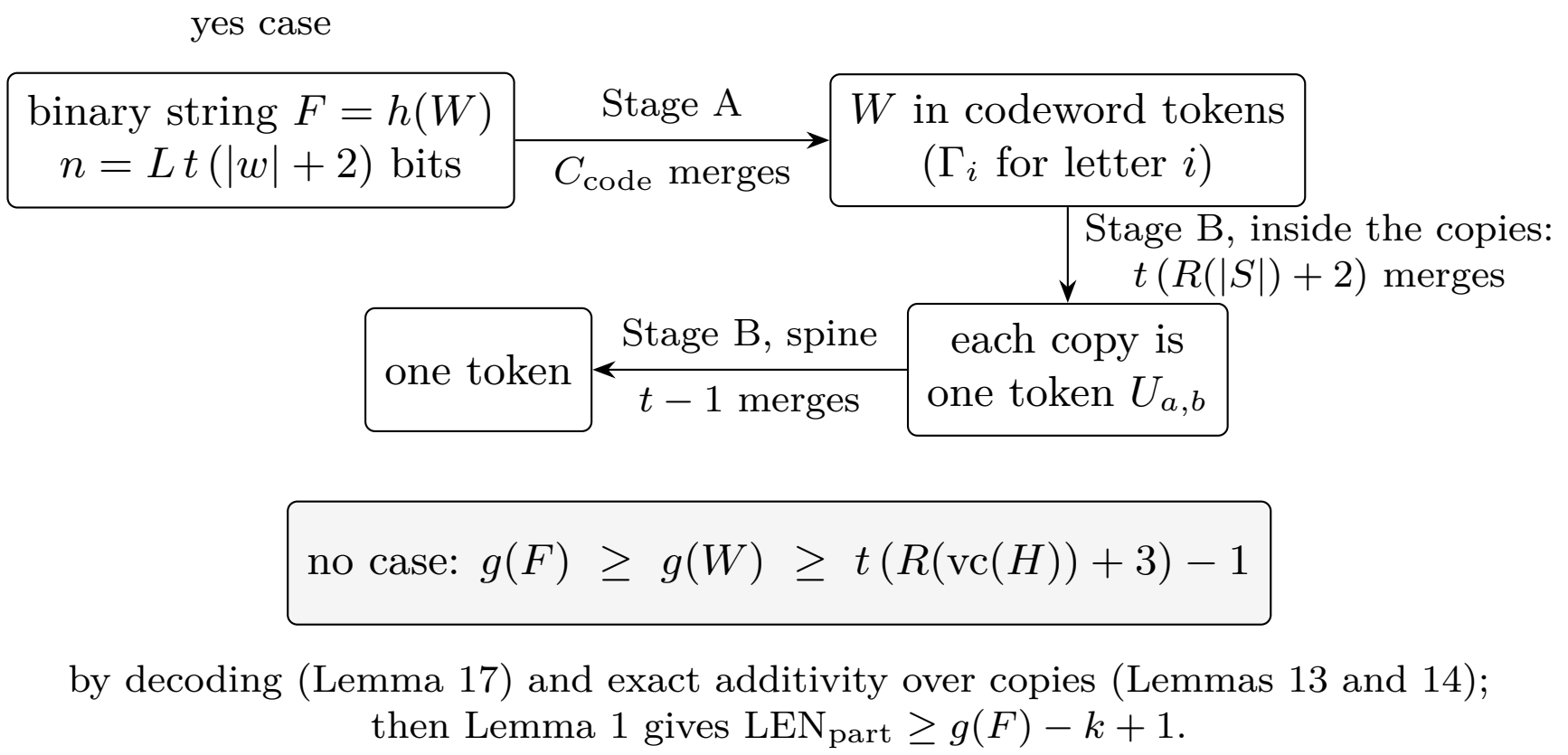
\begin{figure}[!htb]
\centering
\begin{tikzpicture}[>=Stealth, box/.style={draw,rounded corners=2pt,align=center,inner sep=4pt,font=\small,minimum height=9mm},
  lab/.style={font=\footnotesize,align=center}]
\node[box] (F) {binary string $F=h(W)$\\ $n=L\,t\,(|w|+2)$ bits};
\node[box,right=2.3cm of F] (W) {$W$ in codeword tokens\\ ($\Gamma_i$ for letter $i$)};
\draw[->] (F) -- node[lab,above]{Stage A} node[lab,below]{$\Ccode$ merges} (W);
\node[box,below=0.7cm of W] (B) {each copy is\\ one token $U_{a,b}$};
\draw[->] (W) -- node[lab,right]{Stage B, inside the copies:\\ $t\,(R(|S|)+2)$ merges} (B);
\node[box,left=2.3cm of B] (O) {one token};
\draw[->] (B) -- node[lab,above]{Stage B, spine} node[lab,below]{$t-1$ merges} (O);
\node[box,below=0.5cm of O,fill=black!4,xshift=2.2cm] (lb) {no case: $g(F)\ \ge\ g(W)\ \ge\ t\,(R(\vc(H))+3)-1$};
\node[lab,below=0.15cm of lb] {by decoding (\cref{lem:decode}) and exact additivity over copies (\cref{lem:directsum,lem:affine});\\
then \cref{lem:merge2slp} gives $\LENp\ge g(F)-k+1$.};
\node[lab,above=0.1cm of F] {yes case};
\end{tikzpicture}
\caption{Plan of the proof of \cref{thm:main}. Top: the yes case, with $S$ a minimal vertex cover. Bottom: the no
case. Stage~B uses $t\,R(|S|)+2t$ merges inside the copies ($R(|S|)$ for the gadget and $2$ for the guards) and the
spine $t-1$, in total $t(R(|S|)+3)-1$. The budget is $k=\Cbar+t(R(\kappa)+3)-1$, so the difference between the two
cases is at least $t\Delta-\Cbar$ with $\Delta=\vc(H)-\kappa$.}\label{fig:pipeline}
\end{figure}

Recall the template from \cref{sec:prelim}: we need a binary string $F$ and a budget $k$ such that $k$ full merges
reduce $F$ to one symbol in the yes case, while $g(F)-k$ is large in the no case. The gadget
$s_H$ has an additive gap of $\Delta=\vc(H)-\kappa$ rules, but it needs many letters. Writing each letter in binary
costs extra rules. The trick, due to Nowicki \cite{nowicki2026binary}, is to pay that cost once while multiplying the
gap. \Cref{fig:pipeline} shows the plan.
\begin{enumerate}
\item \textbf{Amplify} (\cref{sec:binary-construction}). Take $t$ copies of $w=s_H$, each relabelled by a different
  affine map over a prime field, with guard letters around each copy. No adjacent pair occurs in two copies, so
  grammar size adds up exactly over the copies: the gap becomes $t\Delta$.
\item \textbf{Encode.} Write every letter of the amplified string $W$ as a binary codeword of length $L=O(\log n)$.
  The codewords are designed so that any repeated piece of the binary string $F$ can be read back as a piece of $W$.
  Hence $g(F)\ge g(W)$ (\cref{sec:binary-lower}).
\item \textbf{Schedule} (\cref{sec:binary-upper}). In the yes case, Stage~A uses about $q_W\cdot 2d$ full merges to turn
  every codeword into a single token, giving back $W$. Stage~B then runs the schedule of \cref{lem:sched} inside each
  copy and joins everything. The encoding cost of Stage~A depends on the alphabet size $q_W$ of $W$, not on the number
  $t$ of copies, and with $t=p^2$ it is of lower order than $t\Delta$.
\end{enumerate}

\subsection{The construction}\label{sec:binary-construction}

Let $w$ be a word with property (ALT) (adjacent letters differ, and the first letter differs from the last), with
$q$ distinct letters. In our application $w=s_H$.

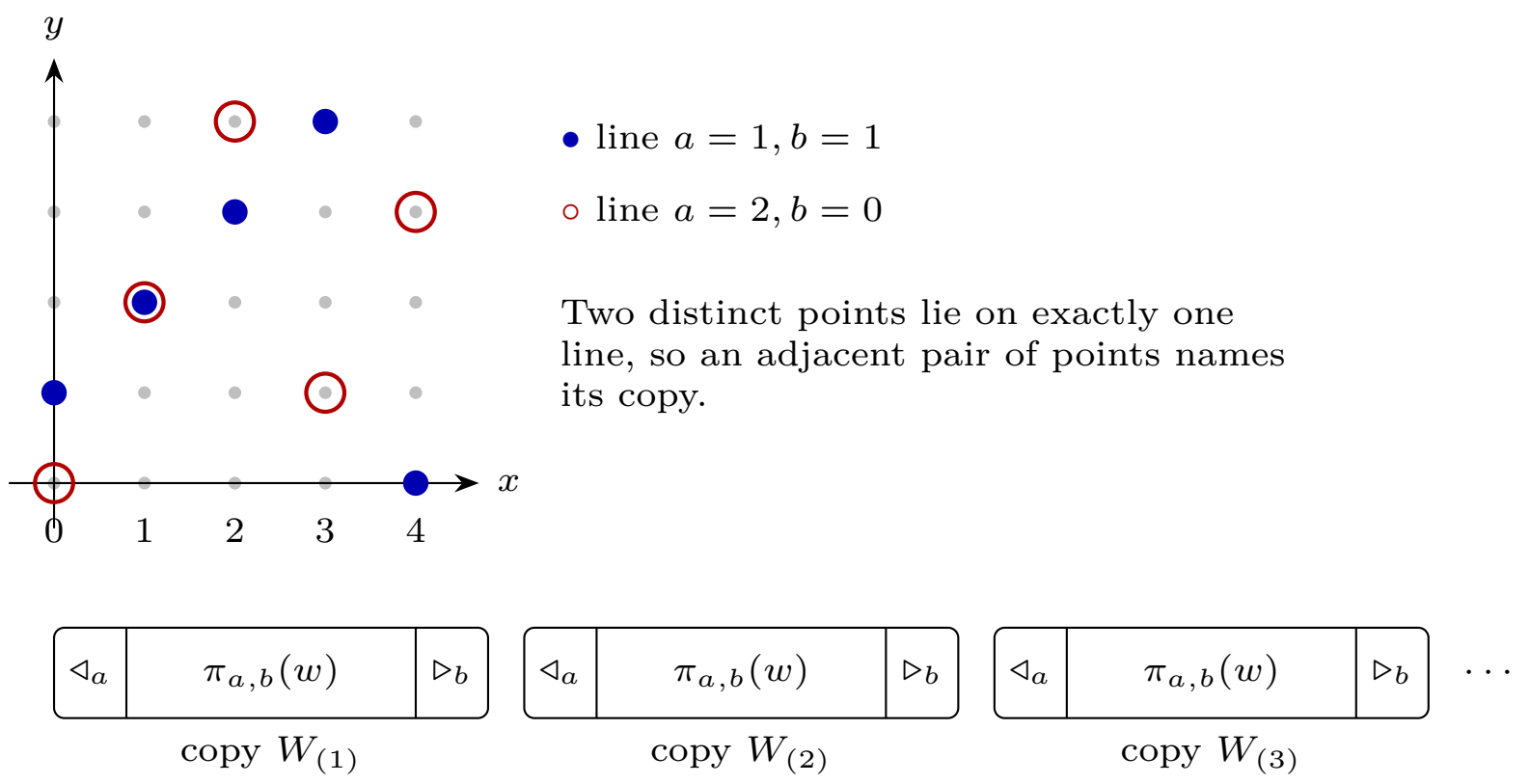
\begin{figure}[!htb]
\centering
\begin{tikzpicture}[scale=0.62,>=Stealth]
\foreach \x in {0,...,4} \foreach \y in {0,...,4} \fill[black!25] (\x,\y) circle (2pt);
\draw[->] (-0.5,0) -- (4.7,0) node[right,font=\scriptsize]{$x$};
\draw[->] (0,-0.5) -- (0,4.7) node[above,font=\scriptsize]{$y$};
\foreach \x in {0,...,4} \node[below,font=\scriptsize] at (\x,-0.2) {\x};
\foreach \x/\y in {0/1,1/2,2/3,3/4,4/0} \fill[blue!70!black] (\x,\y) circle (4pt);
\foreach \x/\y in {0/0,1/2,2/4,3/1,4/3} \draw[red!70!black,thick] (\x,\y) circle (6pt);
\node[font=\scriptsize,anchor=west] at (5.4,3.8) {\textcolor{blue!70!black}{$\bullet$} line $a=1,b=1$};
\node[font=\scriptsize,anchor=west] at (5.4,3.0) {\textcolor{red!70!black}{$\circ$} line $a=2,b=0$};
\node[font=\scriptsize,anchor=west,text width=5.2cm] at (5.4,1.4) {Two distinct points lie on exactly one line, so an
adjacent pair of points names its copy.};
\begin{scope}[yshift=-2.6cm]
\foreach \i/\lab in {0/{(1)},1/{(2)},2/{(3)}} {
  \draw[rounded corners=2pt] (\i*5.2,0) rectangle ++(4.8,1);
  \node[font=\scriptsize] at (\i*5.2+0.4,0.5) {$\gl{a}$};
  \node[font=\scriptsize] at (\i*5.2+2.4,0.5) {$\pi_{a,b}(w)$};
  \node[font=\scriptsize] at (\i*5.2+4.4,0.5) {$\gr{b}$};
  \draw (\i*5.2+0.8,0) -- ++(0,1); \draw (\i*5.2+4.0,0) -- ++(0,1);
  \node[font=\scriptsize,below] at (\i*5.2+2.4,0) {copy $W_{\lab}$};
}
\node[font=\scriptsize] at (15.9,0.5) {$\cdots$};
\end{scope}
\end{tikzpicture}
\caption{Affine amplification, with $p=5$ for illustration. Top: the letters of two copies are points on two
different lines of $\mathbb F_5^2$. Bottom: the copies are concatenated with guards, $W=W_{(1)}W_{(2)}\cdots W_{(t)}$.
Inside a copy every adjacent pair names the copy; between copies the pair $\gr b\gl{a'}$ occurs only once. No adjacent pair is shared by two
copies, which makes grammar size exactly additive (\cref{lem:directsum,lem:affine}).}\label{fig:affine}
\end{figure}

\paragraph{Affine copies.} Choose a prime $p\ge q$ and an injection of the letters of $w$ into the field
$\mathbb F_p$ that sends the first letter of $w$ to $0$ and the last to $1$. (This needs $p\ge2$ and uses the second
half of (ALT).) We identify each letter with its field element. For $a,b\in\mathbb F_p$ let
\[
\pi_{a,b}(x)=(x,\,ax+b),
\]
a ``point'' letter on the line $y=ax+b$. Let $\gl a$ and $\gr b$ be fresh \emph{guard} letters, one for each $a$ and
one for each $b$. The copy indexed by $(a,b)$ is
\[
W_{a,b}=\gl a\;\pi_{a,b}(w)\;\gr b
\]
(\cref{fig:affine}). Two distinct points lie on exactly one line, so an adjacent pair of point letters names its copy.
The guards make the copy boundaries unique as well: without them, the last letter of one copy and the first letter of
the next could form a pair that repeats.
List the pairs $(a,b)$ in row-major order (first by $a$, then by $b$) and let $W=W_{(1)}W_{(2)}\cdots W_{(t)}$ be the
concatenation of the first $t\le p^2$ copies. Let $q_W$ be the number of distinct letters of $W$; then $q_W\le p(q+2)$
(at most $pq$ points and $2p$ guards).

\paragraph{Anchored binary code.} Number the $q_W$ letters of $W$ as $0,\dots,q_W-1$ and let $d=\max(1,\lceil\log_2 q_W\rceil)$.
Encode one bit by a pair: $\eta(0)=\str{10}$ and $\eta(1)=\str{11}$. For a letter $i$ with $d$-bit binary
representation $i_1\cdots i_d$ let $\bar\eta(i)=\eta(i_1)\cdots\eta(i_d)$, a string of $2d$ bits. The codeword of $i$ is
\[
\chi_i=\str{00}\;\bar\eta(i)\;\bar\eta(i)\;\str1,\qquad\text{of length } L=4d+3 .
\]
The \emph{anchor} of a codeword is its bit at offset $2d+2$ (counting from $0$), the first bit of the second copy of
$\bar\eta(i)$. Let $h$ be the morphism that maps letter $i$ to $\chi_i$, and let $F=h(W)$. Then
\[
n=|F|=L\cdot t\cdot(|w|+2).
\]
For example, with $d=4$ the letter $i=5$ has bits $0101$, so
\[
\bar\eta(5)=\str{10\,11\,10\,11},\qquad \chi_5=\str{00\,10111011\,10111011\,1}\quad(L=19).
\]

Three features matter. The marker $\str{00}$ occurs only at codeword starts and fixes the \emph{phase} (the position
modulo $L$). The identifier is
written twice, so a window containing two anchors sees a complete identifier at each end; and every potential $\str0$
after the marker is flanked by $\str1$s, which keeps full merges in step with the pairs (\cref{lem:stageA}).

\subsection{Lower bound: binary encoding does not shrink the grammar}\label{sec:binary-lower}

This subsection states, in our notation, the amplification and decoding lemmas of Nowicki
\cite[Lemmas 3.2, 3.4, 4.1, 4.3, 5.1, 5.3]{nowicki2026binary}. The ideas are Nowicki's; complete proofs are in
\cref{app:binary}, so the argument does not rely on the preprint.

\paragraph{Intuition.} A grammar saves rules only by reusing pieces that occur at least twice. If every adjacent
pair either stays inside one block or occurs only once, then no reusable piece crosses a block boundary, and the
grammar must pay for each block separately (\cref{lem:directsum}). The affine copies have exactly this property
(\cref{lem:affine}). For the binary code, a reusable piece of $F$ that contains two anchors can be read as a piece of
$W$, because equal binary windows have the same phase (\cref{lem:sync}); so a grammar for $F$ can be converted into a
grammar for $W$ that is no larger (\cref{lem:decode}). Nonterminals that contain fewer than two anchors are not
translated at all: they contribute at most one letter of $W$ to the rule above them, so each rule of the binary grammar
yields at most one rule over the letters of $W$.

\begin{definition}
A factorisation $T=T_1\cdots T_r$ is \emph{two-separated} if every adjacent pair $xy$ of $T$ either (a) has all its
occurrences inside one fixed factor $T_i$, or (b) occurs exactly once and crosses a boundary between factors.
\end{definition}

\begin{lemma}\label{lem:twosep-rep}
In a two-separated factorisation, every factor $u$ of $T$ with $|u|\ge2$ that occurs at least twice has all its
occurrences inside one fixed $T_i$.
\end{lemma}

\begin{lemma}[direct sum]\label{lem:directsum}
If $T=T_1\cdots T_r$ is two-separated, then $g(T)\ge\sum_ig(T_i)+r-1$, with $g(\text{letter})=0$.
\end{lemma}

Equality also holds, by concatenating smallest SLPs, but we only need the inequality.

\begin{lemma}[affine separation]\label{lem:affine}
The factorisation $W=W_{(1)}\cdots W_{(t)}$ into copies is two-separated.
\end{lemma}

\begin{lemma}\label{lem:gW}
$g(W)\ge t\,(g(w)+3)-1$. (Equality holds, by joining smallest SLPs for the copies and the guards with $2t+t-1$ rules.)
\end{lemma}

\begin{lemma}[synchronisation]\label{lem:sync}
Let $z$ be a word over the $q_W$ letters (for example $z=W$, so that $h(z)=F$). In $h(z)$ the pair $\str{00}$ occurs
exactly at codeword starts. Suppose two equal factors of $h(z)$ occupy intervals
$I$ and $I'$, and $I$ contains at least two anchors. Then $I$ and $I'$ start at the same offset modulo $L$, and the
sequences of letters whose anchors lie in $I$ and in $I'$ are equal.
\end{lemma}

\begin{lemma}[reverse decoding]\label{lem:decode}
For every word $z$ over the $q_W$ letters with $|z|\ge2$, $g(z)\le g(h(z))$.
\end{lemma}

\begin{proposition}[lower bound]\label{prop:lower}
$g(F)\ge g(W)\ge t(g(w)+3)-1$. For $w=s_H$ this is $t\,(R(\vc(H))+3)-1$.
\end{proposition}

\begin{proof}
\Cref{lem:decode} with $z=W$, then \cref{lem:gW} and \cref{cor:gsH}.
\end{proof}

\subsection{Upper bound: an explicit schedule of full merges}\label{sec:binary-upper}

This is the new part of the proof. In the yes case we must reduce $F$ to a single token using only full merges, and
only about $\Cbar+t(R(\kappa)+3)$ of them. Partial merges could follow any grammar (\cref{lem:realise}); full merges
cannot skip unwanted occurrences, so we must check every merge.

\begin{figure}[!htb]
\centering\small
\renewcommand{\arraystretch}{1.2}
\begin{tabular}{@{}l@{\quad}l@{}}
\toprule
after & codeword $\chi_5$ ($d=4$), followed by the start of the next codeword\\
\midrule
start & $\str0\,\str0\ \ \str1\,\str0\ \ \str1\,\str1\ \ \str1\,\str0\ \ \str1\,\str1\ \ \str1\,\str0\ \ \str1\,\str1\ \ \str1\,\str0\ \ \str1\,\str1\ \ \str1\ \mid\ \str0\,\str0\,\cdots$\\
1: $(\str0,\str0)\to \Omega$ & $\Omega\ \ \str1\,\str0\ \ \str1\,\str1\ \ \str1\,\str0\ \ \str1\,\str1\ \ \str1\,\str0\ \ \str1\,\str1\ \ \str1\,\str0\ \ \str1\,\str1\ \ \str1\ \mid\ \Omega\cdots$\\
2: $(\str1,\str1)\to \beta_1$ & $\Omega\ \ \str1\,\str0\ \ \beta_1\ \ \str1\,\str0\ \ \beta_1\ \ \str1\,\str0\ \ \beta_1\ \ \str1\,\str0\ \ \beta_1\ \ \str1\ \mid\ \Omega\cdots$\\
3: $(\str1,\str0)\to \beta_0$ & $\Omega\ \ \beta_0\ \ \beta_1\ \ \beta_0\ \ \beta_1\ \ \beta_0\ \ \beta_1\ \ \beta_0\ \ \beta_1\ \ \str1\ \mid\ \Omega\cdots$\qquad ($=\tau_5$, $2d+2=10$ tokens)\\
4: trie, depths $2..10$ & $P_{(\Omega,\beta_0)}\ \ \beta_1\cdots$\ \ $\to$\ \ $P_{(\Omega,\beta_0,\beta_1)}\cdots$\ \ $\to\ \cdots\ \to$\ \ $\Gamma_5$\\
\bottomrule
\end{tabular}

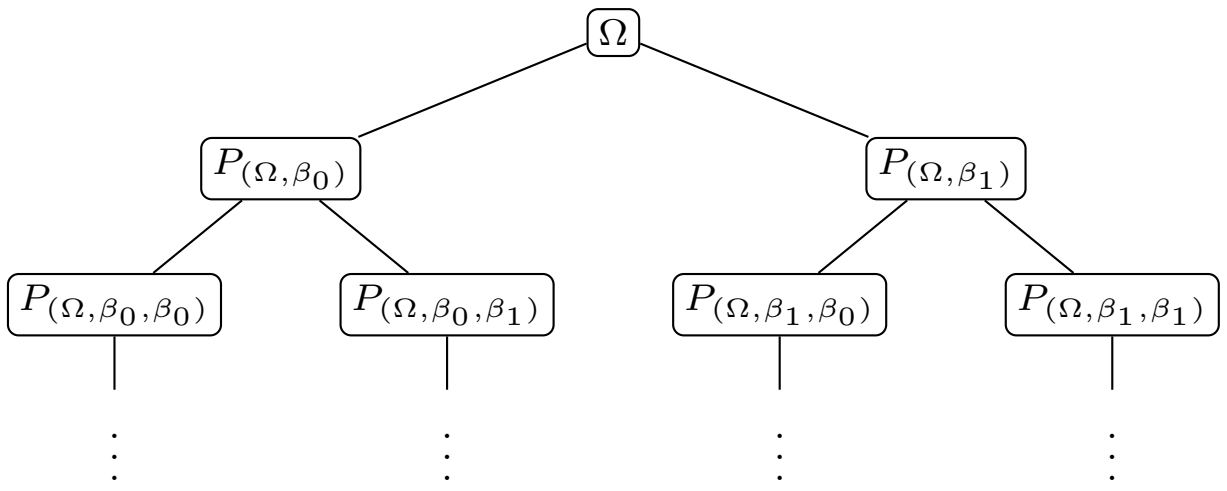
\vspace{6pt}
\begin{tikzpicture}[>=Stealth,level distance=9mm,level 1/.style={sibling distance=44mm},level 2/.style={sibling distance=22mm},
  every node/.style={draw,rounded corners=2pt,font=\scriptsize,inner sep=2pt}]
\node {$\Omega$}
  child {node {$P_{(\Omega,\beta_0)}$}
     child {node {$P_{(\Omega,\beta_0,\beta_0)}$} child {node[draw=none] {$\vdots$}}}
     child {node {$P_{(\Omega,\beta_0,\beta_1)}$} child {node[draw=none] {$\vdots$}}}}
  child {node {$P_{(\Omega,\beta_1)}$}
     child {node {$P_{(\Omega,\beta_1,\beta_0)}$} child {node[draw=none] {$\vdots$}}}
     child {node {$P_{(\Omega,\beta_1,\beta_1)}$} child {node[draw=none] {$\vdots$}}}};
\node[draw=none,align=left,text width=5.6cm,anchor=west] at (4.9,-1.4) {one merge per trie node of depth $\ge2$;
only prefixes that occur are built, so $\Ccode\le3+(2d+1)q_W$. A token $P_\varphi$ only ever sits at the front of a
codeword, so each trie merge fires exactly where intended.};
\end{tikzpicture}
\caption{Stage~A (\cref{lem:stageA}) on one codeword. Top: the three bit-level full merges. In step~2 the run of raw
$\str1$s after a raw $\str0$ has odd length, and the left-to-right scan pairs exactly the code pairs $\str{11}$; the
leftover $\str1$ belongs to the next $\str{10}$. Bottom: the trie of codeword prefixes.}\label{fig:stageA}
\end{figure}

\paragraph{Intuition for Stage A.} We first want every codeword to become one token. Merge $\str{00}$ first: it marks
codeword starts and nothing else. Next merge $\str{11}$, then $\str{10}$. A full merge scans left to right, so a run
of raw $\str1$s is paired up from its left end. Every such run starts at a pair boundary of the code, so the scan
pairs exactly the code pairs $\eta(1)=\str{11}$, and the leftover $\str1$s are the first halves of $\eta(0)=\str{10}$
or the final $\str1$. After these three merges each codeword is a sequence of $2d+2$ tokens. Then we glue tokens from
the left along a trie of codeword prefixes. The token at the front of a codeword is never confused with any other
token, so each trie merge fires exactly at the intended codewords. \Cref{fig:stageA} traces one codeword.

Let $\tau_i=(\Omega,\beta_{i_1},\dots,\beta_{i_d},\beta_{i_1},\dots,\beta_{i_d},\str1)$. This is the token form of $\chi_i$ after the
first three merges below. Define
\[
\Ccode:=3+\sum_{j=2}^{2d+2}\bigl|\{\text{prefixes of length $j$ of the }\tau_i:\ 0\le i<q_W\}\bigr|\ \le\ 3+(2d+1)q_W=:\Cbar .
\]

\begin{lemma}[Stage A: codewords by full merges]\label{lem:stageA}
The following $\Ccode$ full merges turn $F$ into the sequence $W$ in which each letter $i$ is replaced by a token
$\Gamma_i$, with $\Gamma_i\ne\Gamma_j$ for $i\ne j$.
\begin{enumerate}
\item $(\str0,\str0)\to \Omega$;\quad 2.\ $(\str1,\str1)\to \beta_1$;\quad 3.\ $(\str1,\str0)\to \beta_0$;
\setcounter{enumi}{3}
\item for $j=2,\dots,2d+2$, and for each distinct length-$j$ prefix $\varphi=\varphi'\theta$ of the $\tau_i$:
  $(P_{\varphi'},\theta)\to P_\varphi$, where $P_{(\Omega)}=\Omega$. Finally $\Gamma_i:=P_{\tau_i}$.
\end{enumerate}
\end{lemma}

\paragraph{Intuition for Stage B.} After Stage A we have $W$ written in tokens $\Gamma_i$. We now run the schedule of
\cref{lem:sched} copy by copy. A merge of two original tokens could in principle also fire in another copy; but every
pair of point tokens it merges is an adjacent pair of $\pi_{a,b}(s_H)$, and by \cref{lem:affine} such a pair occurs
only in its own copy. Tokens created while processing a copy occur only in that copy. So the copies do not interact.

\begin{lemma}[Stage B: the copies by full merges]\label{lem:stageB}
Let $w=s_H$, and let $S$ be a minimal vertex cover. After Stage A, the following $t\,(R(|S|)+3)-1$ full merges reduce
the sequence to one token.
\begin{itemize}
\item For each copy $(a,b)$ in turn: the schedule $\sched(S)$ of \cref{lem:sched}, with each source letter $x$
  relabelled to $\Gamma_{\pi_{a,b}(x)}$ and fresh tokens for the new symbols. This leaves one token $\Psi_{a,b}$ between the
  guards.
\item Then, before the next copy, $(\Gamma_{\gl a},\Psi_{a,b})\to \Psi'_{a,b}$ and $(\Psi'_{a,b},\Gamma_{\gr b})\to U_{a,b}$.
  (The spine of $\sched(S)$ joins the symbols of the copy between its guards; it never touches a guard.)
\item Finally the spine $(U_{(1)},U_{(2)})\to U_{\le2}$, $(U_{\le2},U_{(3)})\to U_{\le3}$, and so on ($t-1$ merges).
\end{itemize}
\end{lemma}

Both lemmas are proved by tracing every merge; the proofs are in \cref{app:binary-upper}.

\begin{proposition}[yes case]\label{prop:upper}
If $H$ has no isolated vertices and $\vc(H)\le\kappa$, then $\LENf\bigl(F,\ \Cbar+t(R(\kappa)+3)-1\bigr)=1$, and
hence also $\LENp=1$.
\end{proposition}

\begin{proof}
A minimum cover is minimal and $R(\vc(H))\le R(\kappa)$. \Cref{lem:stageA,lem:stageB} use
$\Ccode+t(R(\vc(H))+3)-1$ merges, which is at most the budget; pad with merges of absent pairs. A full merge is a
partial merge, so $\LENp=1$ too.
\end{proof}

\begin{example}[sizes for $K_2$]\label{ex:k2bin}
For $H=K_2$ we have $q=8$ and the smallest admissible prime is $p=11$. Our simulation of the schedule
(\cref{app:repro}) gives, for $t=1$: $q_W=10$, $d=4$, $L=19$, $n=380$ bits, $\Ccode=73$, and $73+(15+3)-1=90$ full
merges reduce $F$ to one token. For all $t=p^2=121$ copies: $q_W=110$, $d=7$, $L=31$, $n=75{,}020$ bits,
$\Ccode=1{,}103$, and $3{,}280$ merges reduce $F$ to one token. The Stage-A cost $1{,}103$ is paid once, while the
per-copy cost $18$ is paid $121$ times.
\end{example}

\subsection{Proof of the main theorem}\label{sec:binary-final}

The reduction maps $(H,\kappa)$ to $(F,k)$ with
\[
k=\Cbar+t\,(R(\kappa)+3)-1,\qquad \Cbar=3+(2d+1)q_W .
\]
It is computable in time polynomial in $n$; the prime $p$ is found by trial division. Write $\Delta=\vc(H)-\kappa$.
\begin{itemize}
\item \textbf{Yes} ($\Delta\le0$): $\LENf(F,k)=\LENp(F,k)=1$ by \cref{prop:upper}.
\item \textbf{No} ($\Delta\ge1$): by \cref{lem:merge2slp} and \cref{prop:lower},
  \[
  \LENf\ \ge\ \LENp\ \ge\ t\,(R(\kappa)+\Delta+3)-1-k+1\ =\ t\Delta-\Cbar+1 .
  \]
\end{itemize}
It remains to choose $p$ and $t$ so that $t\Delta-\Cbar$ is a large fraction of $n$. \Cref{app:params} does this.
For part (i) it takes $p\approx|V|^\gamma$ for a large constant $\gamma$ and $t=p^2$, so that the gap $\Delta\ge1$ given by
NP-hardness, multiplied by $t$, is $n^{1-O(1/\gamma)}$. For part (ii) it takes $p=O(|V|\log|V|)$ and $t=p^2$; then
$\Delta=\Omega(|V|)$ by \cref{fact:cubic}, $t\ge\Cbar+1$, and the gap $t\Delta-\Cbar$ is at least $c_\star n/\log n$, the
$\log n$ being the codeword length.

\paragraph{Datasets and the tokeniser of Kastreva et al.}

\begin{corollary}\label{cor:kastreva}
Unless $\mathrm P=\mathrm{NP}$, binary bottom-up tokenisation with the total-length objective \cite{kastreva2026bounded}
is not in APX, already for one-string datasets: it cannot be approximated within $c_\star n/\log n$.
\end{corollary}

The only difference from full merges is that Kastreva et al.\ identify tokens by the strings they spell; the proof
(\cref{app:binary-upper}) checks that every token of our schedule spells a different string.

\begin{remark}[several strings]\label{rem:lambda}
Padding $F$ with $\lambda-1$ single-bit strings gives datasets (multisets) of $\lambda$ strings, of total length $n$, on
which the ratio $c\,n/(\lambda\log n)$ cannot be beaten, with $c=c_\star/2$ for $\lambda\le n/2$: every merge leaves a
single bit as one token, so the objective is $\LENf(F,k)+\lambda-1$, and $n=|F|+\lambda-1\le2|F|$. This is only the
one-string result in disguise. Every dataset of $\lambda$ strings has optimum at least $\lambda$, so doing nothing is an
$(n/\lambda)$-approximation, and the remark says nothing about datasets of many short distinct strings, such as the word
tables on which BPE is trained (\cref{sec:limits}). The same holds for partial merges.
\end{remark}

\paragraph{Other tokenisers.} The same reduction also covers OPE-seg (vocabulary extracted from $k$ merges, optimal
segmentation) when each merge combines two tokens that already exist, as in the model of Kozma and Voderholzer and
under the reading of \cref{rem:readings}. Both directions are immediate from what we proved. In the yes case, the final token of our
schedule spells $F$, so $F$ is in the extracted vocabulary and costs $1$. In the no case, if $F$ is segmented into
$\ell$ vocabulary tokens, then each token of length at least $2$ has a rule $X\to YZ$ with $Y,Z$ tokens, and the argument
of \cref{lem:merge2slp} gives $g(F)\le k+\ell-1$. If instead merges may combine arbitrary strings, one merge of (all but
the last letter, the last letter) puts a single string into the vocabulary, and the one-string problem is trivial. The
direct tokeniser is not covered (\cref{sec:limits}).

\subsection{\texorpdfstring{Why the ratio is $n/\log n$ and not $n$}{Why the ratio is n/log n and not n}}\label{sec:binary-limits}

The loss is the codeword length $L=\Theta(\log n)$. This route cannot avoid
it: over a fixed alphabet $g=O(n/\log n)$ \cite{kieffer2000grammar}, so any bound of the form ``optimum $1$ versus
$g-k+1$'' gives at most a ratio $O(n/\log n)$. \Cref{sec:gap} shows that, for Karp reductions, this is not an artefact of the method. \Cref{app:binary-polylog} explains why the route of
part (i), using NP-hardness alone, cannot reach $n/\mathrm{polylog}(n)$.

\section{Unary alphabets: strong NP-completeness}\label{sec:unary}

Over the alphabet $\{\str a\}$ a string $\str a^\ell$ is just its length $\ell\ge1$, and a token is a positive integer.
Unary inputs model runs of one character, and repeated strings are natural: BPE is trained on a word-frequency
table, that is, a multiset. Kastreva et al.\ listed two open decision problems here \cite[Section~6]{kastreva2026bounded}:
for unary OPE-seg tokenisation only weak NP-hardness was known, and unary standard bottom-up tokenisation had no hardness
result at all. We show that both are strongly NP-complete on datasets with repeated strings
(\cref{thm:unary,thm:unaryfull}); \cref{sec:limits} discusses what these results do not cover. We also give a closed
formula for a single unary string under full merges (\cref{lem:single}). This section is
self-contained and uses its own local symbols; letters from \cref{tab:notation} may have a different meaning here.

\paragraph{The two tokenisers.} A dataset $\mathcal D$ is a finite multiset of positive integers, each $\ell$ written out
as the string $\str a^\ell$. Datasets may repeat strings, as in Kastreva et al. A merge $(x,y)$ creates the token $x+y$.
\begin{itemize}
\item \emph{OPE-seg tokenisation} \cite[Section~5.2]{kastreva2026bounded}. A \emph{merge sequence} of length $k$ is
  $\langle(x_1,y_1),\dots,(x_k,y_k)\rangle$ with $x_j,y_j\in\mathcal V_{j-1}$, where $\mathcal V_0=\{1\}$ and
  $\mathcal V_j=\mathcal V_{j-1}\cup\{x_j+y_j\}$. Its \emph{vocabulary} is $\mathcal V=\mathcal V_k$; both parts of a
  merge must already exist (see \cref{rem:readings}). Each string is then segmented optimally with $\mathcal V$, at cost
  \[
  \cost_{\mathcal V}(\ell)=\min\{r:\ \ell=v_1+\dots+v_r,\ v_i\in\mathcal V\},
  \]
  which exists because $1\in\mathcal V$.
\item \emph{Standard bottom-up tokenisation}, that is, full merges \cite[Equation~(2)]{kastreva2026bounded}. The merges are
  applied in order to the token string $1^\ell$; each merge $(x,y)$ scans left to right and replaces every
  non-overlapping occurrence of the adjacent pair $x,y$ by the single token $x+y$. Write $\mathrm{tok}^\uparrow_{\mathbf
  m}(\ell)$ for the resulting token string.
\end{itemize}
The decision problems ask, given $\mathcal D$, a budget $k$ and a threshold $\delta$, whether some merge sequence of
length $k$ has total cost $\sum_{\ell\in\mathcal D}\cost_{\mathcal V}(\ell)\le\delta$, respectively total length
$\sum_{\ell\in\mathcal D}|\mathrm{tok}^\uparrow_{\mathbf m}(\ell)|\le\delta$. In both problems ``exactly $k$'' and ``at
most $k$'' merges are equivalent, by monotonicity: appending a merge (for instance $(1,1)$, which is always reachable)
only enlarges the vocabulary, which never raises a segmentation cost, and an appended full merge only shortens strings.

A problem with numbers in its input is \emph{strongly} NP-hard if it stays NP-hard when all numbers are bounded by a
polynomial in the input size, equivalently when they are written in unary \cite{garey1978strong}. Kastreva et
al.\ proved that unary OPE-seg tokenisation is NP-hard by a reduction from addition sequences (addition chains for a set of targets), which are
known to be NP-hard only for binary-encoded targets \cite{downey1981addition}; written out in unary, those targets are exponentially long. So
their result is weak NP-hardness.

\subsection{OPE-seg tokenisation}

\begin{theorem}[unary OPE-seg]\label{thm:unary}
Under the reachable reading (both parts of a merge are existing tokens), unary OPE-seg tokenisation is strongly
NP-complete: it is NP-complete even when the dataset, a multiset with repeated strings, is given as explicit unary
strings.
\end{theorem}

\emph{Membership in NP.} Let the input have total size $\|\mathcal D\|$, so $\max\mathcal D\le\|\mathcal D\|$.
Tokens larger than $\max\mathcal D$ are useless and cannot be parts of smaller tokens, so we may assume every token is
at most $\max\mathcal D$, and at most $\max\mathcal D$ merges are not padding. The certificate is this list of merges.
Costs are computed by the change-making dynamic program in time $O(\max\mathcal D\cdot|\mathcal V|)$.

Hardness is a reduction from vertex cover, which is strongly NP-hard because it has no numbers.

\subsubsection*{The construction}

\paragraph{Intuition.} We want an edge string to be cheap only if one of its endpoints was ``bought''. Every vertex
$v$ gets a number $\alpha_v$, and we ask for a ``cover string'' $\alpha_v+M$ for each vertex and an ``edge string''
$\alpha_u+\alpha_v+M$ for each edge, where $M$ is a large power of two. Buying the token $\alpha_v+M$ makes the cover
string cost $1$ and every edge string at $v$ cost $2$. Two problems must be solved. First, the numbers $\alpha_v$
and $M$ must themselves be tokens, and making a number a token costs merges. We add a \emph{scaffold}: many copies
of every number needed to build them, so that the optimum is forced to build the scaffold, at a known cost. Second,
the scaffold must not create shortcuts. We make all small scaffold numbers powers of two, make all differences of
edge lengths multiples of $3$, and choose the vertex numbers from a Sidon set (all pairwise sums distinct, \cref{lem:sidon}), so that $\theta_u+\theta_v$
determines the edge.

\paragraph{Local notation.} In this section $H=(V,E)$ is a graph with $V=\{1,\dots,|V|\}$, $|V|\ge2$ and $|E|\ge1$,
and $\kappa$ is the vertex-cover threshold. If $\kappa\ge|V|$ the answer is yes and we output a fixed yes-instance; so
assume $0\le\kappa<|V|$. Steps:
\begin{enumerate}
\item Let $\varpi$ be the smallest prime with $\varpi>|V|$; then $3\le\varpi\le2|V|$.
\item For $v=1,\dots,|V|$ let $\xi_v=2\varpi v+(v^2\bmod\varpi)$ \cite{erdos1941sidon}, and $\theta_v=3\xi_v$.
\item Let $\beta$ be the least integer with $2^{\beta-2}>\max_v\theta_v$, and put $U=2^\beta$ and $M=2^{\beta+7}=128U$.
\item For each vertex $v$ let $\alpha_v=9U+\theta_v$. Let $b_1>b_2>\dots>b_r$ be the positions of the $1$-bits of
  $\theta_v$, and let $\phi_{v,i}=U+2^{b_1}+\dots+2^{b_i}$ for $i=1,\dots,r$, the prefix sums. So $\phi_{v,r}=U+\theta_v$.
\item The \emph{scaffold} is $\mathcal S=\mathcal P\cup\Pi\cup\mathcal A$ with $\mathcal P=\{2^i:0\le i\le\beta+7\}$,
  $\Pi=\{\phi_{v,i}\}$ (a set) and $\mathcal A=\{\alpha_v\}$.
\item With $\omega=|V|+|E|+1$, the dataset is the multiset union $\mathcal D=D_1+D_2+D_3$:
  \begin{itemize}
  \item $D_1$: $\omega$ copies of every $z\in\mathcal S\setminus\{1\}$;
  \item $D_2$: one \emph{cover string} $\alpha_v+M$ for every vertex $v$;
  \item $D_3$: one \emph{edge string} $\ell_e=\alpha_u+\alpha_v+M$ for every edge $e=\{u,v\}$.
  \end{itemize}
\item Budget $k=|\mathcal S|-1+\kappa$ and threshold $\delta=\omega(|\mathcal S|-1)+2|V|+2|E|-\kappa$.
\end{enumerate}

\begin{figure}[!htb]
\centering
\begin{tikzpicture}[>=Stealth,font=\scriptsize]
\draw[->] (0,0) -- (15.6,0);
\foreach \x/\l in {0/1,2.2/$U$,4.2/$2U$,5.2/$4U$,6.2/$8U$,8.6/$16U$,9.6/$32U$,10.6/$64U$,11.6/$M{=}128U$} {
  \draw (\x,0.12) -- (\x,-0.12); \node[below] at (\x,-0.12) {\l};}
\draw[very thick,blue!70!black] (0,0.25) -- (2.0,0.25); \node[above,align=center] at (1.0,0.3) {powers of $2$\\ only};
\draw[very thick,blue!70!black] (2.2,0.25) -- (2.75,0.25); \node[above,align=center] at (2.55,0.3) {$\Pi$\\ $[U,\frac54U)$};
\draw[very thick,blue!70!black] (7.0,0.25) -- (7.4,0.25); \node[above,align=center] at (7.2,0.3) {$\mathcal A$\\ $[9U,\frac{37}4U)$};
\draw[very thick,dashed,red!70!black] (12.5,0.25) -- (12.9,0.25); \node[above,align=center] at (12.7,0.3) {cover\\ $M{+}[9U,\frac{37}4U)$};
\draw[very thick,dashed,red!70!black] (14.4,0.25) -- (14.9,0.25); \node[above,align=center] at (14.65,0.3) {edge\\ $M{+}[18U,\frac{37}2U)$};
\end{tikzpicture}
\caption{Where the numbers of the unary OPE-seg reduction lie (schematic, not to scale). Solid bars: the scaffold
$\mathcal S$ (powers of two, prefix sums $\Pi$, vertex numbers $\mathcal A$), each element repeated $\omega$ times in
$D_1$. Dashed bars: the cover strings $\alpha_v+M$ and the edge strings $\ell_e$. The only scaffold element above $M/2$
is $M$ itself, and $\mathcal S$ has no element in $[18U,\frac{37}2U)$, so an edge string is never a sum of two scaffold
elements (\cref{lem:Lfacts}).}\label{fig:unary}
\end{figure}
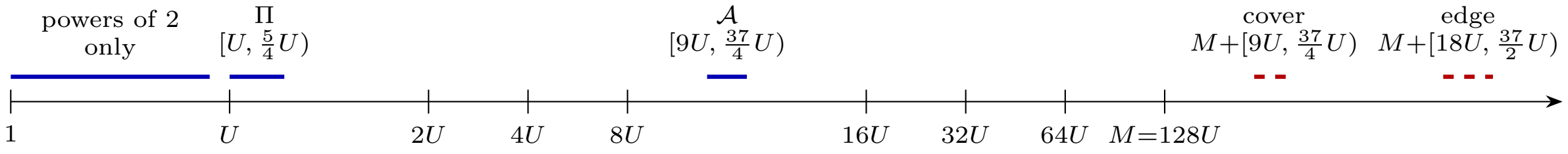

\paragraph{Size.} $\max_v\theta_v<3(2\varpi|V|+\varpi)=O(|V|^2)$, so $U\le8\max_v\theta_v=O(|V|^2)$ and every length is
less than $2M=O(|V|^2)$. Also $|\mathcal S|\le(\beta+8)+|V|\beta+|V|=O(|V|\log|V|)$. The unary input has total length
$O((|V|+|E|)\cdot|V|\log|V|\cdot|V|^2)$, which is polynomial, and the construction runs in polynomial time. \Cref{fig:unary} shows where the numbers lie.

\begin{example}[$K_2$]\label{ex:unaryk2}
For a single edge: $\varpi=3$, $\xi=(7,13)$, $\theta=(21,39)$, $\beta=8$, $U=256$, $M=32768$, $\alpha=(2325,2343)$. The
prefix sums are $272,276,277$ for $\theta_1=21=10101_2$ and $288,292,294,295$ for $\theta_2=39=100111_2$. So
$|\mathcal S|=16+7+2=25$ and $\omega=4$. The cover strings are $35093$ and $35111$, and the edge string is $37436$.
Then $k=24+\kappa$ and $\delta=102-\kappa$. An exhaustive search over all sets of extra buildable tokens, given the scaffold (\cref{app:repro}), finds minimum total cost $103>102$
for $\kappa=0$ and $101=\delta$ for $\kappa=1$, the latter with the single extra token $35093=\alpha_1+M$.
\end{example}

\subsubsection*{Why it works}

The arithmetic facts behind the construction (\cref{lem:sidon,lem:ranges,lem:chain,lem:Lfacts}) and the correctness proof are in \cref{app:unary}. The proof has two
directions. A vertex cover $C$ gives a merge sequence that builds the scaffold and then the cover tokens $\alpha_v+M$
for $v\in C$. Conversely, the weight $\omega$ forces the whole scaffold into the vocabulary. That uses all
but $\kappa$ merges. The remaining $\kappa$ extra tokens must save $\kappa+|E|$ units on cover and edge strings. A
counting argument shows that every extra token is a cover token or an edge token, and that every edge string costs
exactly $2$ or less. The arithmetic facts then show that the only way to write an edge length as a sum of two
tokens is (cover token of an endpoint) + (scaffold element).

\paragraph{Which model.} Kastreva et al.\ do not say whether both parts of a merge must already be tokens.
\Cref{thm:unary} uses the \emph{reachable} reading, in which they must; this matches Kozma and Voderholzer, and it is
the reading under which their own weak NP-hardness proof goes through. Under the literal reading (any pair of strings)
unary OPE-seg coincides with unary direct tokenisation, which they already proved strongly NP-complete. So unary OPE-seg
tokenisation is strongly NP-complete under either reading. \Cref{rem:readings} in \cref{app:unary} gives the details.

\subsection{Standard bottom-up tokenisation (full merges)}\label{sec:unaryfull}

Full merges are rigid: a merge acts on every occurrence of its pair, and the first useful merge must be $(1,1)$. One
might expect this rigidity to make the unary problem easy. For a single string it does (\cref{lem:single}). For datasets
it does not (\cref{thm:unaryfull}).

\paragraph{Local notation.} $\mathrm{ord}_2(x)$ is the exponent of $2$ in $x$. For integers $x\ge1$, $i\ge0$ let
\[
\mathrm{pc}_i(x)=\lfloor x/2^i\rfloor+\mathrm{popcount}(x\bmod2^i),
\]
the least number of powers of two, each at most $2^i$, that sum to $x$ ($\mathrm{popcount}$ counts $1$-bits). Note $\mathrm{pc}_0(x)=x$ and
$\mathrm{pc}_{i+1}(x)=\mathrm{pc}_i(\lfloor x/2\rfloor)+(x\bmod2)$.

\subsubsection*{Structure of full merges on unary strings}

\paragraph{Intuition.} Merging $(1,1)$ pairs up a run of ones; merging $(2,2)$ pairs up the twos; and so on. Call the
merge $(2^j,2^j)$ applied when the current head token is $2^j$ a \emph{doubling}. Between doublings, every other merge
can only glue together the few leftover tokens at the end of a string. These leftovers have strictly decreasing
$2$-adic valuations, so they are all distinct and each pair occurs at most once (\cref{fig:unarystructure}).

\begin{figure}[!htb]
\centering\small
\renewcommand{\arraystretch}{1.15}
\begin{tabular}{@{}lllll@{}}
\toprule
merge & kind & string & head $(2^j)^r$ & tail $u_1\cdots u_c$\\
\midrule
start & & $1^{13}$ & $1^{13}$ & --\\
$(1,1)$ & doubling & $2^6\,1$ & $2^6$ & $1$\\
$(2,2)$ & doubling & $4^3\,1$ & $4^3$ & $1$\\
$(4,4)$ & doubling & $8\;4\;1$ & $8^1$ & $4\;1$\\
$(8,4)$ & non-doubling & $12\;1$ & $8^0$ & $12\;1$\\
\bottomrule
\end{tabular}
\caption{Full merges on the unary string $\str a^{13}$, written as a token string (tokens are numbers). Doublings halve
the head run; leftovers drop into the tail, whose $2$-adic valuations strictly decrease ($\mathrm{ord}_2$ of $4,1$ is
$2,0$; of $12,1$ it is $2,0$). So tail tokens are distinct, and every non-doubling merge acts at most once per string
(\cref{lem:valuation}). Here $4$ merges leave $2$ tokens, the minimum given by \cref{lem:single}:
$\mathrm{pc}_3(13)+3-4=3+3-4=2$.}\label{fig:unarystructure}
\end{figure}

Fix a merge sequence. Keep a counter $j$, initially $0$, and increase it at every merge equal to $(2^j,2^j)$ for the
current $j$ (a doubling). All other merges are \emph{non-doubling}.

The proofs of \cref{lem:valuation,lem:potential,lem:single} are in \cref{app:unary}.

\begin{lemma}[structure]\label{lem:valuation}
At every time, every string of the dataset has the form $(2^j)^r\,u_1u_2\cdots u_c$ with $r\ge0$ and
$j>\mathrm{ord}_2(u_1)>\mathrm{ord}_2(u_2)>\dots>\mathrm{ord}_2(u_c)\ge0$. In particular $c\le j$, the $u_i$ are pairwise
distinct, and none of them is divisible by $2^j$.
\end{lemma}

So on each string every non-doubling merge acts at most once, and a doubling acts only on the head.

\begin{lemma}[potential]\label{lem:potential}
For a string $\ell$, let $\iota$ be the number of non-doubling merges that have acted on it so far. At every time, for
every $j'\ge j$: $\mathrm{pc}_{j'-j}(r)+c\ \ge\ \mathrm{pc}_{j'}(\ell)-\iota$, where the string is
$(2^j)^ru_1\cdots u_c$.
\end{lemma}

\begin{lemma}[one unary string]\label{lem:single}
For $\ell\ge1$ and $k\ge0$, let $J=\min(k,\lfloor\log_2\ell\rfloor)$. The minimum length of $\str a^\ell$ after $k$ full
merges is
\[
\max\bigl(1,\ \mathrm{pc}_J(\ell)+J-k\bigr).
\]
\end{lemma}

This is elementary and may well be folklore; we found no source. It shows that the single-string problem is in P even
when $\ell$ is written in binary.

\subsubsection*{Datasets are hard}

\begin{theorem}[unary full merges]\label{thm:unaryfull}
Unary standard bottom-up tokenisation is strongly NP-complete on datasets (multisets, with repeated strings) of
explicitly written unary strings.
\end{theorem}

\emph{Membership in NP.} Appending merges never increases the total length, so ``exactly $k$'' and ``at most $k$''
merges have the same yes-instances. Each merge that acts lowers the total by at least $1$, so at most $\|\mathcal D\|$
merges act, and the list of acting merges is a certificate checked by simulation in time polynomial in the unary input.
(Membership when lengths are given in binary is not claimed; strong NP-completeness does not need it.)

Hardness is a reduction from balanced biclique \cite[GT24]{garey1979}: given a bipartite graph
$H=(V_1\cup V_2,E)$ and an integer $\kappa\ge1$, are there $S_1\subseteq V_1$ and $S_2\subseteq V_2$ with
$|S_1|=|S_2|=\kappa$ and $S_1\times S_2\subseteq E$? It is NP-complete and has no numbers in its input. We may assume
$|V_1|,|V_2|\ge2$ (add isolated vertices) and $\kappa\le\min(|V_1|,|V_2|)$ (otherwise output a fixed no-instance).

\begin{figure}[!htb]
\centering
\begin{tikzpicture}[font=\scriptsize,>=Stealth]
\fill[blue!12] (0,0) rectangle (2.7,0.6);
\fill[black!10] (2.7,0) rectangle (4.5,0.6);
\fill[red!12] (5.4,0) rectangle (8.1,0.6);
\foreach \b/\x in {8/0,7/1,6/2,5/3,4/4,3/5,2/6,1/7,0/8} {
  \draw (\x*0.9,0) rectangle ++(0.9,0.6); \node at (\x*0.9+0.45,0.3) {\b};}
\draw[decorate,decoration={brace,amplitude=4pt}] (0,0.7) -- (2.7,0.7) node[midway,above=4pt,align=center] {high block: codes of $V_1$\\ (leading bit $8$)};
\draw[decorate,decoration={brace,amplitude=4pt}] (2.7,0.7) -- (4.5,0.7) node[midway,above=4pt] {$\mu_1,\mu_2$};
\draw[decorate,decoration={brace,amplitude=4pt}] (5.4,0.7) -- (8.1,0.7) node[midway,above=4pt,align=center] {low block: codes of $V_2$\\ (leading bit $2$)};
\node[below] at (4.95,0) {unused};
\node[anchor=west,align=left] at (8.5,0.3) {bit positions for $K_{2,2}$};
\begin{scope}[yshift=-1.3cm]
\node[anchor=west] at (0,0) {edge string after the scaffold:};
\node[anchor=west] at (4.6,0) {$c^{(1)}_x\ \ Q_1\ \ Q_2\ \ c^{(2)}_y$};
\node[anchor=west] at (0,-0.6) {$x$ and $y$ chosen:};
\node[anchor=west] at (4.6,-0.6) {$(c^{(1)}_x{+}Q_1)\ \ (Q_2{+}c^{(2)}_y)$};
\node[anchor=west] at (11.0,-0.6) {$2$ tokens};
\node[anchor=west] at (0,-1.1) {only $x$ chosen:};
\node[anchor=west] at (4.6,-1.1) {$(c^{(1)}_x{+}Q_1)\ \ Q_2\ \ c^{(2)}_y$};
\node[anchor=west] at (11.0,-1.1) {$3$ tokens};
\node[anchor=west] at (0,-1.6) {only $y$ chosen:};
\node[anchor=west] at (4.6,-1.6) {$c^{(1)}_x\ \ Q_1\ \ (Q_2{+}c^{(2)}_y)$};
\node[anchor=west] at (11.0,-1.6) {$3$ tokens};
\node[anchor=west] at (0,-2.1) {neither chosen:};
\node[anchor=west] at (4.6,-2.1) {$c^{(1)}_x\ \ (Q_1{+}Q_2)\ \ c^{(2)}_y$ (last merge)};
\node[anchor=west] at (11.0,-2.1) {$3$ tokens};
\end{scope}
\end{tikzpicture}
\caption{The unary full-merge reduction. Top: the bit layout for $K_{2,2}$. Every number is a set of bits; after the
doublings each string is its list of bits from the top down, and the trie merges turn each code into one token. Bottom:
the AND gadget. Choosing $x$ is the merge $(c^{(1)}_x,Q_1)$, choosing $y$ is $(Q_2,c^{(2)}_y)$, and $(Q_1,Q_2)$ comes
last and brings edge strings with neither endpoint chosen from $4$ to $3$ tokens, so an edge string ends with $2$ tokens
only if both endpoints are chosen.}\label{fig:biclique}
\end{figure}
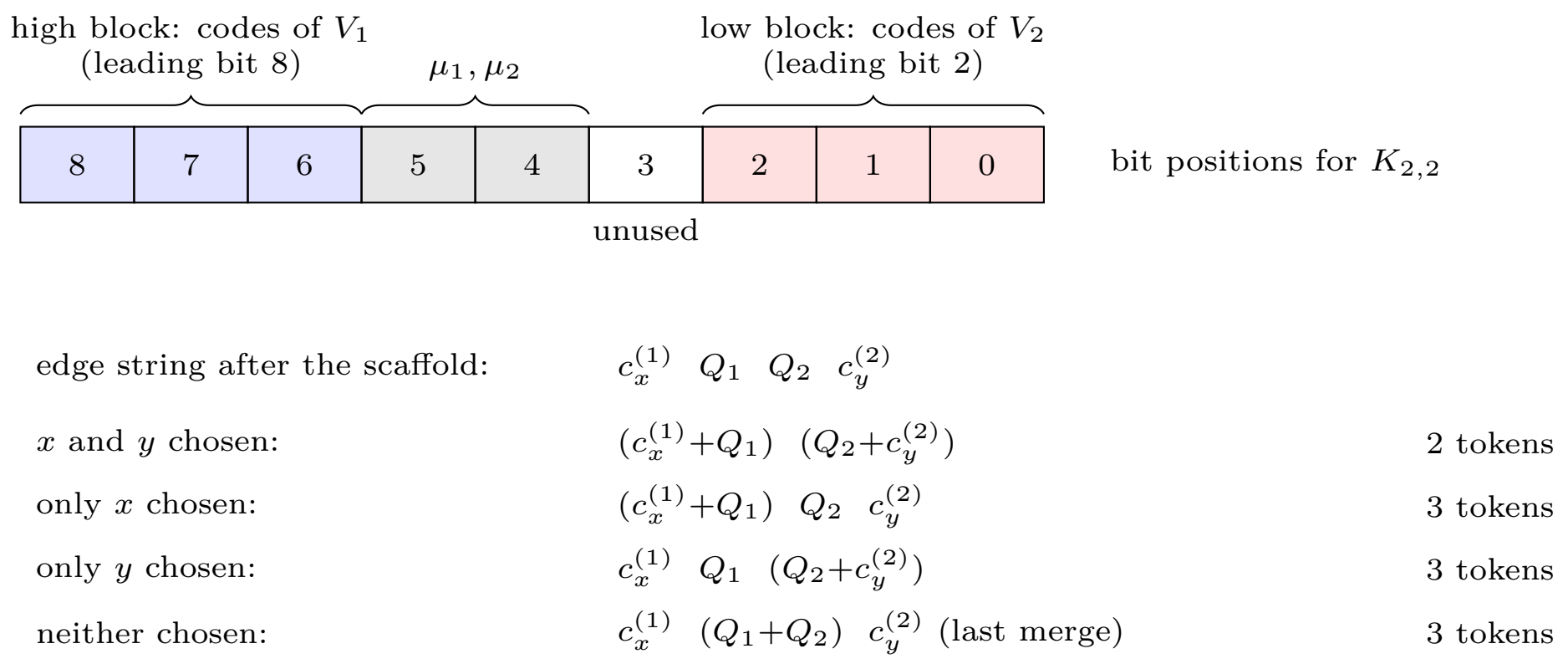

\paragraph{Intuition.} Every number in the construction is a sum of distinct powers of two in disjoint bit positions,
so after enough doublings every string is just its list of bits, from the top down. A vertex $x\in V_1$ gets a binary
code placed in a high block of bits; a vertex $y\in V_2$ gets a code in a low block; and two single bits $Q_1>Q_2$ sit in
between. An edge string reads, after a forced scaffold of merges, $c^{(1)}_x\,Q_1\,Q_2\,c^{(2)}_y$. Choosing $x$ means
merging $(c^{(1)}_x,Q_1)$; choosing $y$ means merging $(Q_2,c^{(2)}_y)$. The merge $(Q_1,Q_2)$, performed \emph{last},
glues the middle of every edge string none of whose endpoints was chosen. So an edge string ends with $2$ tokens if both
endpoints were chosen and with $3$ otherwise: an AND gadget (\cref{fig:biclique}). This is the opposite of the OR
behaviour of vertex cover, and AND is what balanced biclique needs. The forced scaffold makes all codes available as
single tokens at a fixed price. The order of merges matters only for the upper bound. For the lower bound we never reason about the order of full
merges: every output token
lies in the vocabulary $\mathcal V(\mathbf m)=\{1\}\cup\{x+y:(x,y)\in\mathbf m\}$, which has at most $k+1$ elements, so
\begin{equation}\label{eq:LB}
|\mathrm{tok}^\uparrow_{\mathbf m}(\ell)|\ \ge\ \cost_{\mathcal V(\mathbf m)}(\ell).
\end{equation}

\paragraph{Layout.} Let $d_1=\lceil\log_2|V_1|\rceil$ and $d_2=\lceil\log_2|V_2|\rceil$ (both at least $1$). Bits are
numbered from $0$. The low block holds bits $0,\dots,2d_2$, with leading bit $2d_2$; bit $2d_2+1$ is unused; then come
$\mu_2=2d_2+2$ and $\mu_1=2d_2+3$; the high block holds bits $h,\dots,h+2d_1-1$ with $h=2d_2+4$, followed by its leading bit
$h+2d_1$. We identify $V_1$ with $\{0,\dots,|V_1|-1\}$ and $V_2$ with $\{0,\dots,|V_2|-1\}$. For $x\in V_1$ with $d_1$-bit binary
digits $x_1\cdots x_{d_1}$ (most significant first), its \emph{code} is the
decreasing bit list $\mathrm{code}_1(x)=(h+2d_1,\ h+2(d_1-1)+x_1,\ \dots,\ h+2(d_1-i)+x_i,\ \dots)$. Likewise
$\mathrm{code}_2(y)=(2d_2,\ 2(d_2-1)+y_1,\ \dots)$ for $y\in V_2$. Codes of one side have equal length, are distinct, and
none is a proper prefix of another.

\paragraph{Numbers and dataset.} Let $c^{(1)}_x=\sum_{b\in\mathrm{code}_1(x)}2^b$ and $c^{(2)}_y=\sum_{b\in\mathrm{code}_2(y)}2^b$,
$Q_1=2^{\mu_1}$, $Q_2=2^{\mu_2}$, $Q_{12}=Q_1+Q_2$, and $\ell_e=c^{(1)}_x+Q_1+Q_2+c^{(2)}_y$ for an edge $e=(x,y)$. The
scaffold is $\mathcal S=\mathcal P\cup\mathcal T_1\cup\mathcal T_2\cup\{Q_{12}\}$, where $\mathcal P=\{2^i:0\le i\le h+2d_1\}$
and $\mathcal T_i$ is the set of values of the prefixes, of at least two bits, of the codes of side $i$ (a trie; whole
codes included). With $\omega_1=2|E|+1$ and $\omega_0=2\omega_1(|V_1|+|V_2|)+3|E|+1$:
\begin{itemize}
\item $D_1$: $\omega_0$ copies of each $z\in\mathcal S\setminus\{1\}$;
\item $D_2$: $\omega_1$ copies of each \emph{vertex string} $c^{(1)}_x+Q_1$ ($x\in V_1$) and $Q_2+c^{(2)}_y$ ($y\in V_2$);
\item $D_3$: one \emph{edge string} $\ell_e$ per edge.
\end{itemize}
Budget $k=|\mathcal S|-1+2\kappa$ and threshold
$\delta=\omega_0(|\mathcal S|-1)+\omega_1\bigl(2(|V_1|+|V_2|)-2\kappa\bigr)+3|E|-\kappa^2$.

\paragraph{Size.} Every length is below $2^{h+2d_1+1}=2^{2d_1+2d_2+5}<512\,|V_1|^2|V_2|^2$; $|\mathcal S|=O((|V_1|+|V_2|)\log(|V_1|+|V_2|))$;
multiplicities are $O(|E|(|V_1|+|V_2|))$. So the unary instance has polynomial size and is built in polynomial time.

\paragraph{Bit-set view.} All these numbers are sums of distinct powers of two, so we identify each with its set of bits.
Scaffold elements are single bits, prefixes of high codes (containing the high leading bit), prefixes of low codes
(containing $2d_2$), or $\{\mu_1,\mu_2\}$. A vertex string is $\mathrm{code}_1(x)\cup\{\mu_1\}$ or $\{\mu_2\}\cup\mathrm{code}_2(y)$,
and $\ell_e=\mathrm{code}_1(x)\cup\{\mu_1,\mu_2\}\cup\mathrm{code}_2(y)$.

\begin{example}[$K_{2,2}$]\label{ex:k22}
Here $d_1=d_2=1$, the low block is bits $0..2$, bit $3$ is unused, $\mu_2=4$, $\mu_1=5$, and the high block is bits $6,7$
with leading bit $8$. The codes are $(8,6),(8,7)$ on side $1$ and $(2,0),(2,1)$ on side $2$, so
$c^{(1)}\in\{320,384\}$, $c^{(2)}\in\{5,6\}$, $Q_1=32$, $Q_2=16$, $Q_{12}=48$, and the longest edge string is
$384+48+6=438$. The scaffold has $|\mathcal S|=9+2+2+1=14$ elements, and $\omega_1=9$, $\omega_0=85$. For $\kappa=1$:
$k=15$ and $\delta=1170$; for $\kappa=2$: $k=17$ and $\delta=1149$.
\end{example}

\paragraph{The proof in brief.} A balanced biclique $S_1\times S_2$ gives a merge sequence of length $k$: doublings,
the trie merges that build the scaffold, the $2\kappa$ choice merges, and finally $(Q_1,Q_2)$; it reaches total length
exactly $\delta$ (\cref{fig:biclique}). Conversely, the weight $\omega_0$ forces the scaffold into the vocabulary, the
weight $\omega_1$ forces exactly $2\kappa$ choice tokens, and a residue argument modulo $2^{\mu_2}$ shows that an edge
string costs $2$ only if both endpoints were chosen (\cref{lem:edgecost}); the $\kappa^2$ savings then force a
biclique. The full proof is in \cref{app:unary}.

\begin{remark}\label{rem:fullremarks}
\begin{enumerate}\renewcommand{\labelenumi}{(\alph{enumi})}
\item \emph{Readings and budget.} A merge whose pair never occurs has no effect, so the result holds whether or not the
  parts of a merge must be existing tokens. The ($\Leftarrow$) direction uses only $|\mathcal V(\mathbf m)|\le k+1$, and
  the ($\Rightarrow$) sequence is reachable and uses exactly $k$ merges, so ``exactly $k$'' and ``at most $k$'' are both
  covered.
\item \emph{A second proof for OPE-seg.} The ($\Leftarrow$) direction is an OPE-seg lower bound, and the full-merge output of the
  ($\Rightarrow$) sequence is one segmentation into its (reachable) vocabulary. So the same instance also proves
  \cref{thm:unary}, independently of the vertex-cover construction.
\item \emph{Scope.} The weights $\omega_0,\omega_1$ use repeated strings, as Kastreva et al.\ do; see \cref{sec:limits}
  for what is not covered.
\end{enumerate}
\end{remark}

\mypart{Algorithms and the limits of hardness proofs}\label{part:algo}
\section{Approximation algorithms and a barrier for Karp reductions}\label{sec:gap}

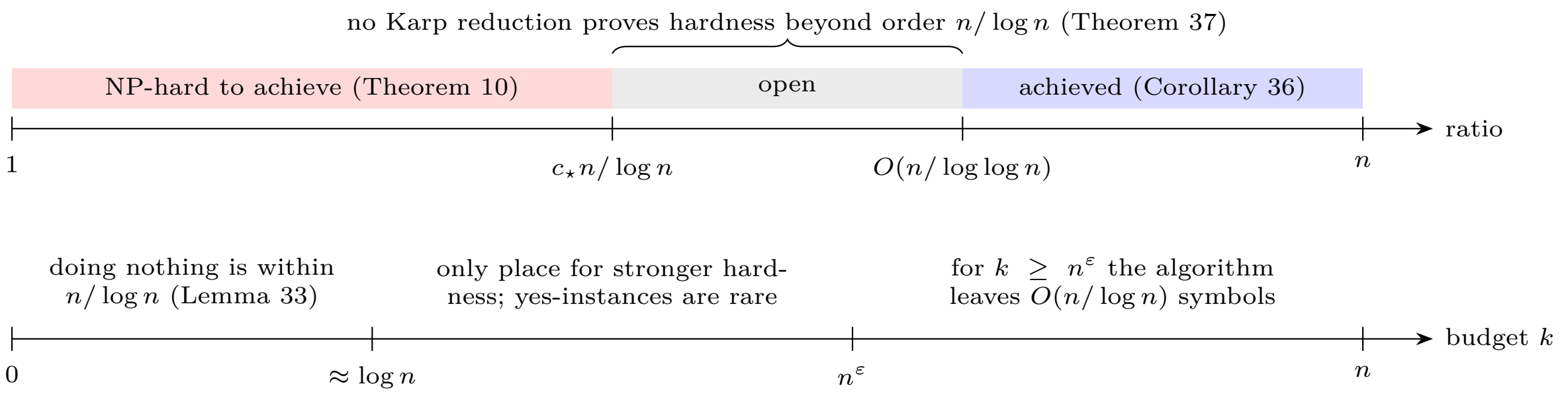
\begin{figure}[!htb]
\centering
\begin{tikzpicture}[>=Stealth,font=\scriptsize]
\draw[->] (0,0) -- (14.2,0) node[right]{ratio};
\foreach \x/\l in {0/$1$,6/{$c_\star n/\log n$},9.5/{$O(n/\log\log n)$},13.5/$n$} {\draw (\x,0.12) -- (\x,-0.12); \node[below] at (\x,-0.15) {\l};}
\fill[red!15] (0,0.2) rectangle (6,0.6); \node at (3,0.4) {NP-hard to achieve (\cref{thm:main})};
\fill[black!8] (6,0.2) rectangle (9.5,0.6); \node at (7.75,0.4) {open};
\fill[blue!15] (9.5,0.2) rectangle (13.5,0.6); \node at (11.5,0.4) {achieved (\cref{cor:algo})};
\draw[decorate,decoration={brace,amplitude=4pt}] (6,0.75) -- (9.5,0.75);
\node[align=center] at (7.75,1.05) {no Karp reduction proves hardness beyond order $n/\log n$ (\cref{thm:barrier})};
\begin{scope}[yshift=-2.1cm]
\draw[->] (0,0) -- (14.2,0) node[right]{budget $k$};
\foreach \x/\l in {0/$0$,3.6/$\approx\log n$,8.4/$n^{\varepsilon}$,13.5/$n$} {\draw (\x,0.12) -- (\x,-0.12); \node[below] at (\x,-0.15) {\l};}
\node[align=center,text width=3.4cm] at (1.8,0.55) {doing nothing is within $n/\log n$ (\cref{lem:lenlower})};
\node[align=center,text width=4.4cm] at (6.0,0.55) {only place for stronger hardness; yes-instances are rare};
\node[align=center,text width=4.6cm] at (11.0,0.55) {for $k\ge n^\varepsilon$ the algorithm leaves $O(n/\log n)$ symbols};
\end{scope}
\end{tikzpicture}
\caption{\Cref{part:algo} for partial merges on binary strings. Top axis: what is known about the best polynomial-time
approximation ratio (schematic). Bottom axis, a separate scale, the budget $k$: why the barrier holds. Small budgets are easy by \cref{lem:lenlower}, and large
budgets by the algorithm. In between, an instance with optimum $o(\log n)$ is determined by fewer than $n^\varepsilon$
merge rules and a short final sequence, so such instances are too rare to be the image of a Karp reduction from SAT
unless $\mathrm{coNP}\subseteq\mathrm{NP/poly}$.}\label{fig:landscape}
\end{figure}

\Cref{thm:main} leaves a gap between $c_\star n/\log n$ and the trivial factor $n$. Here we narrow it from both sides.
For partial merges, a simple algorithm achieves $O(n/\log\log n)$, and no Karp reduction can prove more than
$\Theta(n/\log n)$ unless $\mathrm{coNP}\subseteq\mathrm{NP/poly}$. So, for partial merges and Karp reductions,
\cref{thm:main} is optimal up to the constant, and the remaining gap is between $n/\log n$ and $n/\log\log n$. For full
merges both bounds are weaker by an exponent $\log_32\approx0.63$. \Cref{fig:landscape} summarises the picture. This
part does not use the gadgets of \cref{part:hardness}. A \emph{Karp reduction} is a polynomial-time map from
SAT to instances, the usual form of an NP-hardness proof; $\mathrm{coNP}\subseteq\mathrm{NP/poly}$ would mean that
unsatisfiability has polynomial-size certificates given advice, and it would collapse the polynomial hierarchy. This
section uses local symbols: $\sigma$ is the alphabet size, $N$ the length of a SAT instance, $m$ a block length, $I$ a
number of phases, $\mathcal T$ the set of symbols present at the start of a phase, and $\theta_{\mathrm{yes}},\theta_{\mathrm{no}}$
the thresholds of a gap reduction; logarithms are to base $2$, and the budget $k$ is written in binary in the input.

\begin{lemma}[trivial lower bound]\label{lem:lenlower}
Every symbol present after $k$ partial or full merges spells at most $2^k$ letters. Hence
$\LENf(s,k)\ge\LENp(s,k)\ge\lceil n/2^k\rceil$.
\end{lemma}

\begin{proof}
The symbol created by the $i$-th merge joins two symbols that existed before it; by induction they spell at most
$2^{i-1}$ letters each.
\end{proof}

\paragraph{Upper bounds.} With partial merges, cut the string into blocks of length $m$ and build every block from the
left, one letter at a time, along the trie of block prefixes. This is the all-words idea behind LZ78-style grammars, and
we do not claim it as new.

\begin{theorem}[partial merges]\label{thm:opeupper}
For a string $s$ over $\sigma$ letters and an integer $m\ge1$ with $\sum_{\ell=2}^{m}\sigma^\ell\le k$,
$\LENp(s,k)\le\lceil n/m\rceil$. For binary strings one can take $m=\lfloor\log(k+4)\rfloor-1$, and for $\sigma$ letters
$m=\lfloor\log_\sigma k\rfloor-1$; the $\sigma$-ary choice needs $k\ge\sigma^2$, so that $m\ge1$ (the binary one gives
$m\ge1$ for every $k\ge0$).
\end{theorem}

Full merges cannot respect a block alignment. Instead we work in phases. In each phase, every adjacent pair of old
symbols is merged, and then every old symbol is merged with the new symbol to its right. This halves the length, while
the number $|\mathcal T|$ of distinct symbols grows to at most $3|\mathcal T|^3$ per phase.

\begin{theorem}[full merges, binary]\label{thm:omsupper}
For a binary string $s$ and an integer $I\ge0$ with $\log k\ge1.8\cdot3^I+2$, $\LENf(s,k)\le(n-1)/2^I+1$. Hence, for $\log k>2$,
$\LENf(s,k)\le2n\bigl(1.8/(\log k-2)\bigr)^{\log_32}+1$. Over $\sigma$ letters the same holds with $1.8$ replaced by
$1.8\log\sigma$.
\end{theorem}

The proofs are in \cref{app:gap}. A program written separately from the proofs simulates both constructions with exact merge
semantics (\cref{app:repro}).

\begin{corollary}[the algorithm]\label{cor:algo}
Fix $\sigma$. For $n\ge n_0(\sigma)$, outputting the better of ``do nothing'' and the construction of
\cref{thm:opeupper} is a polynomial-time $O(n\log\sigma/\log\log n)$-approximation of $\LENp$; for binary strings the
construction of \cref{thm:omsupper} gives an $O\bigl(n/(\log\log n)^{\log_32}\bigr)$-approximation of $\LENf$. Moreover,
on inputs with $k\ge n^{\varepsilon}$ the binary partial-merge construction leaves at most $n/(\varepsilon\log n-2)+1$
symbols, the $\sigma$-ary one at most $n\log\sigma/(\varepsilon\log n-2\log\sigma)+1$, and the binary full-merge one at most
$2n\bigl(1.8/(\varepsilon\log n-2)\bigr)^{\log_32}+1$.
\end{corollary}

\begin{proof}
If $2^k\le n/\log n$, doing nothing has ratio at most $n/\lceil n/2^k\rceil\le2^k\le n/\log n$ by \cref{lem:lenlower}.
Otherwise $k>\log n-\log\log n\ge(\log n)/2$ for $n\ge16$, and \cref{thm:opeupper} leaves at most
$n/(\log_\sigma k-2)+1\le n\log\sigma/(\log\log n-1-2\log\sigma)+1$ symbols, against an optimum of at least $1$; the
denominator is positive for $n\ge n_0(\sigma)$. The full-merge case is the same with \cref{thm:omsupper}. The strong
form substitutes $k\ge n^{\varepsilon}$. For polynomial time when $k$ is written in binary, cap $k$ at $n-1$ (later
merges cannot act), output the at most $nm$ merges of the construction explicitly, and describe the remaining identity
merges implicitly.
\end{proof}

\paragraph{The barrier.} A Karp gap reduction maps satisfiable formulas to instances whose optimum is at most a
threshold $\theta_{\mathrm{yes}}(n)$ and unsatisfiable ones to instances whose optimum is at least
$\theta_{\mathrm{no}}(n)$, with ratio $\rho=\theta_{\mathrm{no}}/\theta_{\mathrm{yes}}$. We assume, as usual for gap
problems, that the thresholds are polynomial-time computable and $1\le\theta_{\mathrm{yes}}(n)<\theta_{\mathrm{no}}(n)\le n$.

\begin{theorem}[no Karp reduction beyond $n/\log n$]\label{thm:barrier}
If approximating $\LENp$ on binary strings within a ratio $\rho(n)$ with $\rho(n)\log n/n\to\infty$ is NP-hard by a
Karp gap reduction, then $\mathrm{coNP}\subseteq\mathrm{NP/poly}$, and the polynomial hierarchy collapses to its third
level \cite{yap1983}. The same holds for $\LENf$ on binary strings when $\rho(n)(\log n)^{\log_32}/n\to\infty$, and for
$\LENp$ over $\sigma$ letters when $\rho(n)\log n/(n\log\sigma)\to\infty$.
\end{theorem}

\emph{Proof idea.} By \cref{cor:algo}, a no-instance (optimum at least $\theta_{\mathrm{no}}=\omega(n/\log n)$) must have
$k<n^{\varepsilon}$. Yes-instances with $k<n^{\varepsilon}$ and optimum $\theta_{\mathrm{yes}}=o(\log n)$ are few: such a
string is determined by $k$ merge rules and a short final sequence, so there are at most $2^{O(n^{\varepsilon}\log n)}$
of them. Intuitively, so few instances force many formulas, or tuples of formulas, to share an image, so a polynomial list of
no-instance images, given as advice, lets an NP machine certify unsatisfiability. Formally, a reduction from
SAT to such a sparse set forces $\mathrm{coNP}\subseteq\mathrm{NP/poly}$ by the density
theorem of Buhrman and Hitchcock \cite[Lemmas 3.2 and 3.5]{buhrman2008dense}. The full proof, including the order in
which the running time of the reduction and $\varepsilon$ are chosen, is in \cref{app:gap}. The theorem covers Karp
reductions from any NP-hard problem, but not Turing or randomised reductions, and says nothing about assumptions such as
the exponential-time hypothesis (ETH: 3-SAT on $v$ variables cannot be solved in time $2^{o(v)}$).

\paragraph{Large budgets.} Fix $0<\varepsilon<1$. Over all binary inputs with $k\ge n^{\varepsilon}$, the worst-case
approximability of $\LENp$ is $\Theta_\varepsilon(n/\log n)$ unless $\mathrm P=\mathrm{NP}$: \cref{cor:algo} leaves
$O_\varepsilon(n/\log n)$ symbols on such inputs, and the hard instances of \cref{thm:main} have
$k=\Theta(n/\log n)\ge n^{\varepsilon}$. The worst case is attained at $k=\Theta(n/\log n)$; for budgets
$n^{\varepsilon}\le k\le n^{1-\nu}$ (fixed $\nu>0$) no matching lower bound is known. The only budgets where hardness beyond
$n/\log n$ could hold are $\log n\lesssim k<n^{\varepsilon}$; this is what the barrier exploits.

\paragraph{A conjecture.} We expect the truth to be $\Theta(n/\log\log n)$ for partial merges.

Gap-ETH strengthens ETH: for some constant $\varepsilon>0$, even distinguishing satisfiable 3-SAT formulas on $v$
variables from those in which every assignment violates an $\varepsilon$ fraction of the clauses cannot be done in time
$2^{o(v)}$.

\begin{conjecture}\label{conj:gapeth}
Under Gap-ETH \cite{dinur2016gapeth,manurangsi2016birthday}, for some $c>0$ no polynomial-time algorithm approximates
$\LENp$ or $\LENf$ on binary strings within $c\,n/\log\log n$.
\end{conjecture}

The idea is to take a gap vertex-cover instance on $N_0=\mathrm{polylog}(n)$ vertices, which Gap-ETH makes hard in
polynomial time, encode it as a binary word $w$ of length $\Theta(N_0\log N_0)$ with a constant number of copies, and use
$s=w^M$ with $M=2^j$ and budget $k=g(w)+j$. In the yes case $g(w)$ merges build $w$ and $j$ doublings finish. The missing
piece is a lemma that, in the no case, every choice of $k$ merges cuts each copy of $w$ into $\Omega(N_0)$ pieces, which
would leave $\Omega(n/\log\log n)$ tokens. \Cref{lem:merge2slp} is too weak here, because the grammar size of $w^M$ is
only about $g(w)+j$; the lemma needs a robust per-copy count of vertex-cover witnesses. We have not proved it.

\mypart{Discussion}\label{part:discussion}
\section{Related work and connections}\label{sec:related}

\begin{itemize}
\item \emph{Tokenisation complexity.} Zouhar et al.\ \cite{zouhar2023formal} started the formal analysis of BPE. Kozma
  and Voderholzer \cite{kv2026arxiv} introduced OPE, restated the problem of Zouhar et al.\ as OMS, proved APX-hardness
  of utility over unbounded alphabets (which extends to compressed length), and analysed BPE and EvenOdd. Whittington et al.\ \cite{whittington2025npcomplete}
  proved NP-completeness over unbounded alphabets. Kastreva et al.\ \cite{kastreva2026bounded} proved NP- and APX-hardness
  of binary direct and bottom-up tokenisation, strong NP-completeness of unary direct tokenisation, and weak NP-hardness of
  unary OPE-seg tokenisation.
\item \emph{Grammar compression.} BPE run to exhaustion is Re-Pair \cite{larsson2000repair,navarro2008repair}. Textual
  substitution schemes are NP-hard to optimise \cite{storer1982}, and the smallest grammar problem is NP-hard and APX-hard
  \cite{charikar2005smallest,lehman2002thesis}. Bannai et al.\ \cite{bannai2021revisited} revisit the approximation
  ratios of grammar compressors, including Re-Pair, and relate grammar sizes over arbitrary and binary alphabets.
  Arpe and Reischuk \cite{arpe2006grammar} also relate the approximability of grammar compression over arbitrary and
  binary alphabets. Grammar size is measured either by the total length of the right-hand sides or, as here, by the number of binary rules.
  Casel et al.\ \cite{casel2021fixed} prove NP-completeness and APX-hardness for the right-hand-side measure over fixed
  alphabets of at least $17$ letters. Over a binary alphabet, NP-completeness for both the right-hand-side and the
  binary-rule measure was claimed by Masierak \cite{masierak2026binary} (Nowicki's appendix disputes the threshold used
  there) and by Nowicki \cite{nowicki2026binary}, and for the right-hand-side measure by Chen \cite{chen2026sgp}; none
  of them gives an approximation lower bound. For a different objective, partition-cover (direct) tokenisation, Lim et
  al.\ \cite{lim2025partition} prove NP-hardness and our preprint \ourcite{baliga2026greedtok} gives a constant-factor
  approximation.
\item \emph{Unbounded alphabets.} The $\Theta(n)$ bound of \cref{thm:unbounded} first appeared in the earlier versions (up to
  version~1.2) of our unreviewed preprint on byte-pair encoding \ourcite{baliga2026bpevonetwo}; that preprint now refers to this
  paper for it. It answers the question of Kozma and Voderholzer up to a constant factor over unbounded alphabets, and the
  binary result builds on it. Its gadget follows the vertex-cover
  reduction of Charikar et al.\ \cite{charikar2005smallest}. Nowicki \cite[Section~7.2]{nowicki2026binary} quotes an exact
  formula, due to Kawamoto, I, K\"oppl and Bannai \cite{kawamoto2024rlslp}, for the smallest SLP of a related vertex-cover
  gadget. Combined with \cref{fact:cubic}, that formula would imply APX-hardness of grammar size in binary rules over
  unbounded alphabets, as already noted in \ourcite{baliga2026bpevonetwo}; Masierak
  \cite{masierak2026asi} also states APX-hardness of this measure (as the assembly index), citing the smallest-grammar
  literature. We rely on Nowicki's statement of the formula of \cite{kawamoto2024rlslp} and do not use it in any proof.
\item \emph{What we take from Nowicki.} Nowicki's preprint \cite{nowicki2026binary} gives exact bounds on binary grammar
  size (his Theorem~6.1) and says in its Section~8: ``The proof does not establish a binary approximation lower bound.''
  We use his affine copies with guards and the two-separation direct sum (\cref{lem:twosep-rep,lem:directsum,lem:affine,lem:gW};
  his Lemmas 3.2, 3.4, 4.1, 4.3), and his anchored code with reverse decoding (\cref{lem:sync,lem:decode}; his
  Section~5). Our code $\chi_i=\str{00}\,\bar\eta(i)\,\bar\eta(i)\,\str1$ with $\eta(0)=\str{10}$, $\eta(1)=\str{11}$ is
  identical to his, except that we number the $q_W$ letters actually used instead of his bound $p(q+2)$ and take
  $d\ge1$; we take all $t=p^2$ copies instead of his $t=12p(q+2)+1$, only to fit the larger cost of our Stage~A
  (the ratio does not depend on $t$ once $t$ is well above that cost). His bounds already give an additive gap in binary grammar size, so the partial-merge part of
  \cref{thm:main} follows from them and \cref{lem:merge2slp,lem:realise}; we include it for completeness and make the gap
  explicit. Nowicki's upper bound is a grammar, which full merges cannot follow. \emph{New here} are the full-merge
  schedule (\cref{lem:stageA,lem:stageB}), which uses specific features of the code (the marker, and the $\str1$s around
  every free $\str0$) to keep left-to-right full merges aligned with the pairs, and the consequences for the merge
  problems: \cref{thm:main} for full merges and \cref{cor:kastreva}. Also new are \cref{thm:unary,thm:unaryfull}, the
  barrier of \cref{thm:barrier} and the full-merge bound of \cref{thm:omsupper}; \cref{lem:single} and the upper bound
  of \cref{thm:opeupper} are elementary and may be folklore.
\item \emph{Unreviewed sources.} Nowicki \cite{nowicki2026binary}, Masierak \cite{masierak2026binary}, Chen
  \cite{chen2026sgp} and our own \ourcite{baliga2026greedtok,baliga2026bpe,baliga2026bpevonetwo} are preprints, and \cite{kv2026arxiv} is cited in its extended arXiv
  version. We credit them for ideas but rely on none of them: \cref{sec:warmup} proves in full the result that first
  appeared in \ourcite{baliga2026bpevonetwo}, and \cref{sec:binary-lower} re-proves what we use from \cite{nowicki2026binary}. 
\end{itemize}

\paragraph{Connections: good ratios for tokens saved, none for tokens left.} Positive results for tokenisation
measure the \emph{utility}, the number of symbols saved. Greedy BPE saves at least a third of the optimum for both
partial and full merges \cite{kv2026arxiv}, and at least $0.391$ of it by our preprint \ourcite{baliga2026bpe};
EvenOdd saves half for partial merges \cite[Theorem~4]{kv2026arxiv}; and for partition-cover tokenisation, which is
NP-hard \cite{lim2025partition}, GreedTok achieves a constant fraction of the optimal objective
\ourcite{baliga2026greedtok}. These constant ratios do not contradict \cref{thm:main}. Utility and compressed length add up
to $n$. Our hard instances have budget $k=\Theta(n/\log n)\ge n^{\varepsilon}$, so even in the no case the optimum leaves only $o(n)$
symbols (\cref{cor:algo}): the yes and no cases differ by a vanishing fraction of the symbols saved, which no utility
ratio sees, but by a factor of order $n/\log n$ in the symbols left. So a constant-factor guarantee on the tokens saved
does not give a useful guarantee on the tokens left.

\paragraph{Connections to grammar compression.} The bridge ``merges are grammars'' ties tokenisation to the smallest grammar problem in both
directions: grammar lower bounds give hardness for compressed length (\cref{lem:merge2slp}), and partial merges realise
any grammar (\cref{lem:realise}). The upper bound of \cref{thm:opeupper} is the all-words trie behind LZ78-style
grammars. The $\log n$ in \cref{thm:main} has the same source: over a fixed alphabet every string has a grammar with
$O(n/\log n)$ rules \cite{kieffer2000grammar}, so an ``optimum $1$ versus $g-k+1$'' argument cannot give more. The
barrier of \cref{thm:barrier} is a sparse-set argument in the line of Mahaney's theorem \cite{mahaney1982sparse}, using the
density form of Buhrman and Hitchcock \cite{buhrman2008dense}, whose combinatorial core goes back to Fortnow and Santhanam
\cite{fortnow2011infeasibility}. Similar limits on Karp reductions are known for another compression-like problem, the
minimum circuit size problem \cite{murray2017mcsp,hirahara2016mcsp}. Finally, the same string $F$ shows that the binary grammar size $g$, counted in binary rules,
is APX-hard to compute, with ratio $1+\Theta(\varepsilon_0)$, since $g(F)$ lies between $g(W)$ and $g(W)+\Cbar$ by
\cref{prop:lower,prop:upper,lem:merge2slp} (using equality in \cref{lem:gW}). For the right-hand-side measure, APX-hardness
over alphabets of at least $17$ letters is due to Casel et al.\ \cite{casel2021fixed}; for the binary-rule measure over
unbounded alphabets it follows from the formula of Kawamoto et al.\ (see above). The binary
case also follows from Nowicki's bounds, which he states only for exact hardness, and approximability transfers between
alphabets are studied in \cite{arpe2006grammar,bannai2021revisited}; we claim no novelty for this remark.

\section{Scope, limitations and open problems}\label{sec:limits}

\paragraph{Does the hard regime match practice?} No, and the result should not be read as saying that BPE is bad on
natural text. In our hard instances either $k$ merges shrink the whole string to a single token, or every choice of $k$
merges leaves many tokens. The budget is about $t\,R(\kappa)$ while $n=L\,t\,(|s_H|+2)$, and $R(\kappa)$ is a constant
fraction of $|s_H|$; so $k=\Theta(n/\log n)$. Real tokenisers use $k\ll n$ and never reach one token. What the result does
say is that, unless $\mathrm P=\mathrm{NP}$, no polynomial-time training algorithm, BPE or otherwise, can come with a worst-case guarantee on the final token count that
holds on all inputs and is much better than doing nothing. Over all inputs with $k\ge n^{\varepsilon}$ (fixed
$0<\varepsilon<1$) the worst-case ratio for partial merges is $\Theta(n/\log n)$, attained by inputs with
$k=\Theta(n/\log n)$ (\cref{sec:gap}). It says nothing about budgets $k\le n^{1-\nu}$ (fixed $\nu>0$), such as those of real
tokenisers; that regime is open. Our objective is also the token count of the
training string itself, not of held-out text. Worst-case guarantees are better stated for the tokens saved, where
greedy BPE saves at least a third of the optimum \cite{kv2026arxiv}; the connections paragraph of \cref{sec:related}
explains why such guarantees and our hardness coexist. The constant $c_\star$ is tiny and not explicit, so the result is
asymptotic.

\paragraph{What is not covered.}
\begin{itemize}
\item \emph{Other tokenisers and corpora.} Pre-tokenised corpora (word-frequency tables with many short distinct
  strings, merges never crossing word boundaries) are not covered: \cref{rem:lambda} only pads one hard string, and its
  bound degrades by the number of strings, down to the trivial factor $n/\lambda$. Direct
  tokenisation (choose any $k$ tokens, then segment optimally), the closest model of Unigram- or WordPiece-style
  vocabulary selection, is not covered: for one string, a vocabulary containing that string gives length $1$.
\item \emph{Unary results.} \Cref{thm:unary,thm:unaryfull} need datasets with repeated strings, and
  \cref{thm:unary} assumes that merges combine existing tokens (\cref{rem:readings}). Neither gives approximation
  hardness: the objective is dominated by the forced scaffold. For a single unary string under OPE-seg we do not know the complexity; with
  threshold $1$ it asks for a shortest addition chain.
\item \emph{The barrier.} \Cref{thm:barrier} rules out Karp reductions only, not Turing or randomised reductions or
  hardness under other assumptions, and for full merges it stops at $n/(\log n)^{\log_32}$, not $n/\log n$.
\end{itemize}

\subsection{Open problems}\label{sec:open}

\paragraph{The practical regime.} Is compressed length still hard to approximate when $k\le n^{1-\nu}$, or when the
optimum is at least $n^{\nu}$? Is there an algorithm with ratio $o(n)$ that exploits a small budget?

\paragraph{Closing the gap.} For partial merges over a binary alphabet the best factor lies between $c_\star n/\log n$
(hardness, optimal for Karp reductions by \cref{thm:barrier}) and $O(n/\log\log n)$ (\cref{cor:algo}). Grammar compressors with approximation ratio $O(\log(n/g))$
\cite{charikar2005smallest,rytter2003lz,jez2015recompression} do not obviously help here, because in the window
$\log n\lesssim k<n^{\varepsilon}$ the merge budget leaves no room for a multiplicative loss in grammar size. Closing the gap from
the hardness side needs a stronger assumption, such as Gap-ETH (\cref{conj:gapeth}), or non-Karp reductions. For full
merges the algorithm is weaker, $O(n/(\log\log n)^{\log_32})$; an upper bound $\LENf(s,k)=O(n/\log k)$ would align the two
cases. Whether $n/\mathrm{polylog}(n)$ hardness follows from NP-hardness without the PCP theorem is also open
(\cref{app:binary-polylog}).

\paragraph{Unary tokenisation.} Are unary OPE-seg or unary bottom-up tokenisation hard on datasets of distinct strings,
or hard to approximate? Greedy BPE is not optimal for unary full-merge tokenisation on datasets: an exhaustive search
found lengths $(44,47,51,57)$ with multiplicities $(2,1,1,2)$ and $k=7$, where the optimum is $16$ tokens and BPE, with every
tie-breaking rule, gets $17$ (\cref{app:repro}).

\paragraph{Direct tokenisation.} Kastreva et al.\ ask whether binary direct tokenisation is in APX
\cite[Section~4.2]{kastreva2026bounded}; this half of their question stays open. On datasets the optimum is at least
the number of strings, so a superconstant ratio needs a per-string gap that grows.

\paragraph{Utility.} The question of Kozma and Voderholzer whether some algorithm saves more than half of the optimum
remains open for partial and full merges \cite[Section~6]{kv2026arxiv}.

\section*{Use of AI tools} AI-assisted tools, including Claude (Anthropic), were used throughout the research process to support mathematical reasoning, computational experimentation, proof verification, and manuscript preparation. The authors guided the research direction, directed and evaluated the central arguments, and oversaw the computational and verification workflows. Additional independent AI-assisted checks were employed to assess proof consistency and computational correctness. The authors take full responsibility for the results and conclusions presented.

{\small
\let\oldthebibliography\thebibliography
\renewcommand\thebibliography[1]{\oldthebibliography{#1}\setlength{\itemsep}{1pt}\setlength{\parsep}{0pt}}
\bibliographystyle{plainurl}
\bibliography{refs}}

\appendix
\crefalias{section}{appendix}\crefalias{subsection}{subappendix}
{\small
\section{Proofs for the binary lower bound}\label{app:binary}
These are the proofs of \cref{sec:binary-lower}. They re-prove, in our notation, lemmas of Nowicki \cite{nowicki2026binary}.

\begin{proof}[Proof of \cref{lem:twosep-rep}]
If an occurrence of $u$ crossed a boundary, it would contain the boundary pair $xy$, which is of type (b) and occurs
once. The other occurrence of $u$ would give a second occurrence of $xy$ at a different position, a contradiction.
So all occurrences of $u$ are inside factors. The first pair of $u$ occurs at least twice, so it is of type (a): all
its occurrences lie in one fixed $T_i$. Hence so do all occurrences of $u$.
\end{proof}

\begin{proof}[Proof of \cref{lem:directsum}]
Take a smallest SLP for $T$ and its parse tree. Call a nonterminal \emph{shared} if it labels at least two nodes. Two
such nodes are not nested (acyclicity), so they have disjoint yields, and the value, of length at least $2$, is a
repeated factor. By \cref{lem:twosep-rep} it lies in one fixed factor, which we call its segment. Descendants of shared
nodes are shared, and they have the same segment. Contract every topmost shared node to a leaf. What remains is a full
binary tree whose internal nodes carry pairwise distinct, unshared nonterminals. If it has $\Lambda$ leaves, it has
$\Lambda-1$ internal nodes. Each leaf is a letter or a shared value, so no leaf crosses a boundary, and the leaves
split into consecutive groups of sizes $r_i\ge1$, one group per factor. Let $m_i$ be the number of shared nonterminals
with segment $i$. Then $T_i$ can be derived with $m_i+r_i-1$ rules: the $m_i$ shared rules (closed under descendants)
plus a left comb over its $r_i$ leaves. Hence
\[
g(T)\ \ge\ \sum_im_i+\Lambda-1\ =\ \sum_i(m_i+r_i-1)+r-1\ \ge\ \sum_ig(T_i)+r-1 .\qedhere
\]
\end{proof}

\begin{proof}[Proof of \cref{lem:affine}]
A pair of two point letters $(x,ax+b)(y,ay+b)$ has $x\ne y$ by (ALT). Two distinct points lie on exactly one line, so
the pair determines $(a,b)$, and all its occurrences are in $W_{a,b}$: type (a). The pair $\gl a(0,b)$ determines
$(a,b)$, because the first letter of $w$ is $0$. The pair $(1,a+b)\gr b$ determines $b$ and then $a$, because the last
letter of $w$ is $1$. Both are type (a). The pairs that cross a boundary are $\gr b\gl{a'}$. In row-major order each
ordered pair of guards occurs at most once at a boundary, also in any prefix of the list, and no pair inside a copy
consists of two guards. So boundary pairs are type (b).
\end{proof}

\begin{proof}[Proof of \cref{lem:gW}]
$\pi_{a,b}$ is injective on letters (look at the first coordinate), so $g(\pi_{a,b}(w))=g(w)$. The factorisation
$(\gl a,\pi_{a,b}(w),\gr b)$ of a single copy is two-separated: the guards occur once in the copy, and the pairs of
$\pi_{a,b}(w)$ stay inside it. By \cref{lem:directsum}, $g(W_{a,b})\ge g(w)+2$. By \cref{lem:affine,lem:directsum}
for $W$, $g(W)\ge t(g(w)+2)+t-1$.
\end{proof}

\begin{proof}[Proof of \cref{lem:sync}]
Write $\chi_i=\str0\,\str0\,\str1\,b_1\,\str1\,b_2\cdots\str1\,b_d\,\str1\,b_1\cdots\str1\,b_d\,\str1$. Every bit after
offset $1$ that can be $\str0$ is some $b_j$, which is preceded by $\str1$ and followed by $\str1$ (the first bit of
the next pair, or the final $\str1$). Codewords end in $\str1$. So $\str{00}$ occurs exactly at offsets $0\bmod L$.

Let $I$ contain two consecutive anchors $\alpha$ and $\alpha+L$. The $\str{00}$ that starts the next codeword
occupies positions $\alpha+2d+1$ and $\alpha+2d+2$, strictly between the two anchors, so it lies in $I$. At the same
relative position $I'$ contains $\str{00}$, which is a codeword start. So the phases agree. The identifier read at the
first anchor is the second copy of $\bar\eta(i)$, at positions $[\alpha,\alpha+2d-1]\subset I$ (because
$\alpha+L\in I$). For the last anchor $\alpha'$ in $I$, the first copy of its identifier sits at
$[\alpha'-2d,\alpha'-1]\subset I$ (because $\alpha'-L\in I$). Codewords strictly between them lie fully inside $I$.
Equal bits at equal phase therefore give equal identifiers.
\end{proof}

\begin{proof}[Proof of \cref{lem:decode}]
Take any binary SLP for $h(z)$. Call a nonterminal $X$ \emph{macro} if one of its occurrences in the parse tree
contains at least two anchors. By \cref{lem:sync} every occurrence then does, with the same phase and the same
sequence $\mathrm{dec}(X)$ of anchored letters. Process the macro rules $X\to YZ$ in topological order, using one fixed
occurrence of $X$. In that occurrence each child contributes either
\begin{itemize}
\item a nonterminal $\hat Y$ with $\val(\hat Y)=\mathrm{dec}(Y)$, if the child is macro;
\item the single anchored letter, if the child contains exactly one anchor (the letter is read off this fixed
  occurrence of $X$; the child's value alone need not determine it);
\item nothing, if the child contains no anchor.
\end{itemize}
If both children contribute, create one rule $\hat X\to(\cdot)(\cdot)$. Otherwise let $\hat X$ be an alias of the
contributing child, at no cost. (Both cannot be empty, since $X$ has two anchors.) By induction $\val(\hat X)=\mathrm{dec}(X)$.
The root contains all anchors and $|z|\ge2$, so it is macro and decodes to $z$. The new SLP has at most as many rules
as there are macro rules.
\end{proof}

\subsection{\texorpdfstring{Why part (i) does not give $n/\mathrm{polylog}(n)$}{Why part (i) does not give n/polylog(n)}}\label{app:binary-polylog}
If the source word has an additive gap $\Delta$ and
length $|w|$, then $t$ copies give a gap of at most $t\Delta$ while $n=L\,t\,(|w|+2)$, so the ratio is at most
$n\Delta/(L(|w|+2))$. With only NP-hardness, $\Delta=1$, and a ratio within a polylogarithmic factor of $n$ would need
$|w|=\mathrm{polylog}(n)$, forcing $n$ to be exponential in $|w|$. A ratio $n/\mathrm{polylog}(n)$ needs $\Delta=\Omega(|w|)$, which is what the PCP-based \cref{fact:cubic} provides.
Whether NP-hardness alone suffices by another route is open.

\section{Proofs for the binary upper bound and the main theorem}\label{app:binary-upper}
These are the case checks of \cref{sec:binary-upper,sec:binary-final}.

\begin{proof}[Proof of \cref{lem:stageA}]
\emph{Step 1.} By \cref{lem:sync}, $\str{00}$ occurs exactly at codeword starts. Starts are $L\ge7$ apart and
$\str{000}$ never occurs, so exactly the markers are replaced.

\emph{Step 2.} After step 1 a codeword reads $\Omega\,\str1\,b_1\,\str1\,b_2\cdots\str1\,b_d\,\str1\,b_1\cdots\str1\,b_d\,\str1$.
Every maximal run of raw $\str1$s begins right after $\Omega$ or right after a raw $\str0$ (some $b_j=\str0$), so at a
pair boundary: either at the first bit of a pair $\str1b_j$, or at the final $\str1$. The run consists of zero or more
pairs $\str{11}$, followed either by the first bit of a pair $\str{10}$ or by the final $\str1$, and then by $\Omega$ or
the end of the string. A left-to-right,
non-overlapping replacement inside a run that starts on a pair boundary replaces exactly the pairs
$\eta(1)=\str{11}$. A run never crosses $\Omega$.

\emph{Step 3.} The remaining raw $\str1$s are first bits of pairs $\eta(0)=\str{10}$ (followed by a raw $\str0$) and
final $\str1$s (followed by $\Omega$ or the end). The remaining raw $\str0$s are second bits of $\eta(0)$. So exactly the
pairs $\eta(0)$ are replaced, and each codeword now reads $\tau_i$.

\emph{Step 4.} Invariant before depth $j$: codeword $i$ reads $P_{\tau_i[1..j-1]},\tau_i[j],\dots$, and a token
$P_\varphi$ occurs only as the first token of codewords with prefix $\varphi$. This holds at $j=2$, because $\Omega$ occurs
only at codeword starts. A merge $(P_{\varphi'},\theta)$ therefore occurs exactly at the codewords with prefix
$\varphi'\theta$. These occurrences do not overlap, because $P_{\varphi'}\ne\theta$: here $\theta\in\{\beta_0,\beta_1,\str1\}$, while
$P_{\varphi'}$ is $\Omega$ or a merged token. The merges at depth $j$ touch disjoint sets of codewords, and all codewords
have the same token length $2d+2$. So the invariant holds at $j+1$. At $j=2d+2$ each codeword is one token
$P_{\tau_i}$, and these are distinct because distinct letters have distinct $\tau_i$.

\emph{Count.} Three merges, plus one per distinct prefix of length at least $2$.
\end{proof}

\begin{proof}[Proof of \cref{lem:stageB}]
We claim that while copy $(a,b)$ is processed, every occurrence of a merged pair lies inside block $(a,b)$. A merged
pair either contains a token created by this copy's merges, which by induction occurs only in block $(a,b)$, or
consists of two raw Stage-A tokens. Two raw tokens adjacent in the current sequence were adjacent in $W$, because
merges replace contiguous pairs. A pair of point tokens merged by $\sched(S)$ is an adjacent pair of
$\pi_{a,b}(s_H)$, and by \cref{lem:affine} it occurs in $W$ only in block $(a,b)$. Also $\sched(S)$ never merges a
pair containing a guard. So, restricted to block $(a,b)$, the run is exactly the run of $\sched(S)$ on $s_H$ under an
injective relabelling, and by \cref{lem:sched} it leaves $\Gamma_{\gl a}\,\Psi_{a,b}\,\Gamma_{\gr b}$. The two guard merges
then occur once each, because $\Psi_{a,b}$ and $\Psi'_{a,b}$ are specific to this copy. The tokens $U_{a,b}$ are pairwise distinct, so
each spine pair occurs once. The count is $t\cdot R(|S|)+2t+(t-1)=t(R(|S|)+3)-1$.
\end{proof}

\subsection{Choice of parameters for the main theorem}\label{app:params}
We continue the proof of \cref{thm:main} from \cref{sec:binary-final}, with the reduction $(H,\kappa)\mapsto(F,k)$
defined there.

\paragraph{Decision rule.} Suppose an algorithm has ratio $\rho(n)$. For graphs below a constant size, where the
asymptotic estimates below may fail, solve vertex cover by brute force; inputs with $\kappa<|V|/2$ in part (ii) are
answered ``no'' (\cref{sec:prelim}). Otherwise run the
algorithm on $(F,k)$ and answer yes if and only if the length it returns is at most $\rho(n)$. Yes-instances give at
most $\rho(n)\cdot1$. No-instances give at least the optimum, and we show below that the optimum exceeds $\rho(n)$, a
strict gap.

\paragraph{Part (i): $n^{1-\varepsilon}$, from NP-hardness only.} Delete isolated vertices (this does not change
$\vc$) and assume $|E|\ge1$ and $|V|\ge3$. Then $q=1+3|V|+|E|\le2|V|^2$. Fix an integer $\gamma\ge3$ and let $p$ be
the least prime with $p\ge|V|^\gamma$; by Bertrand's postulate $p\le2|V|^\gamma$. Also $p\ge|V|^\gamma\ge2|V|^2\ge q$,
as the injection into $\mathbb F_p$ requires. Take $t=p^2$. Then $q_W\le p(q+2)$ and $d=O(\log|V|)$, so
$\Cbar=O(p\,|V|^2\log|V|)\le p^2/2$ for large $|V|$; this is where $\gamma\ge3$ is needed. Hence:
\begin{itemize}
\item no-instances have $\LENf\ge\LENp\ge p^2-p^2/2+1=p^2/2+1$, and yes-instances have length $1$;
\item $n=(4d+3)\,p^2\,(|s_H|+2)\le p^2\,|V|^{2+o(1)}$, so $p^2/2\ge n/|V|^{2+o(1)}$;
\item $n\ge p^2\ge|V|^{2\gamma}$, so $|V|^{2+o(1)}\le n^{(1+o(1))/\gamma}$, and the gap is at least
  $n^{1-(1+o(1))/\gamma}$.
\end{itemize}
Given $\varepsilon>0$, choose $\gamma>\max(3,1/\varepsilon)$. Then for large $|V|$ the no-instance optimum exceeds
$n^{1-\varepsilon}$, so an $n^{1-\varepsilon}$-approximation would decide vertex cover.

\paragraph{Part (ii): $c_\star n/\log n$, from \cref{fact:cubic}.} Now $H$ is cubic, yes means $\vc(H)\le\kappa$, no
means $\vc(H)\ge(1+\varepsilon_0)\kappa$, and $\kappa\ge|V|/2$; so in the no case $\Delta$ is an integer with
$\Delta\ge\varepsilon_0\kappa\ge\varepsilon_0|V|/2$. Here $q=4.5|V|+1$ and $|s_H|=15|V|$. Let $p$ be the least prime with
\begin{equation}\label{eq:pchoice}
p^2\ \ge\ 4+\bigl(2\lceil\log_2(p(q+2))\rceil+1\bigr)\,p\,(q+2).
\end{equation}
The right side is at most $4+(2\log_2p+2\log_2(q+2)+3)\,p\,(q+2)$. So \eqref{eq:pchoice} holds for every $p\ge p_0$,
where $p_0=c_1q\log_2q$ for a suitable absolute constant $c_1$ and $q$ large; by Bertrand's postulate there is a prime in
$[p_0,2p_0]$, so $p\le2p_0=O(|V|\log|V|)$. Also \eqref{eq:pchoice} gives $p^2\ge3p(q+2)$, hence $p>q$. Take $t=p^2$. Since
$q_W\le p(q+2)$ and $d\le\lceil\log_2(p(q+2))\rceil$, \eqref{eq:pchoice} implies $t\ge\Cbar+1$.
\begin{itemize}
\item No-instances have $\LENf\ge\LENp\ge t\Delta-\Cbar+1\ge t(\Delta-1)+2\ge t(\varepsilon_0|V|/2-1)$; yes-instances
  have length $1$.
\item $n=(4d+3)\,t\,(15|V|+2)$, so the gap is at least $n(\varepsilon_0|V|/2-1)/\bigl((4d+3)(15|V|+2)\bigr)$. Since
  $n\ge2q_W$, we have $d\le\log_2n$, and the gap is at least $c_\star n/\log n$ for a suitable constant $c_\star>0$,
  proportional to $\varepsilon_0$, and large $|V|$.
\end{itemize}
This proves \cref{thm:main}. \hfill$\square$

\begin{proof}[Proof of \cref{cor:kastreva}]
Kastreva et al.'s bottom-up tokeniser applies each merge to all occurrences, left to right, as full merges do. The
only difference is that their tokens are identified by the strings they spell, not by fresh names. In the yes case
this changes nothing, because every token created by our schedule spells a different string:
\begin{itemize}
\item Stage-A tokens spell distinct codeword prefixes, of length less than $L$, or whole codewords.
\item Per-copy Stage-B tokens spell $h(z)$ for a factor $z$ of one copy with $|z|\ge2$, so they have length at least
  $2L$. Tokens of different copies spell different strings, because $z$ contains an adjacent pair confined to its copy
  (\cref{lem:affine}). Within a copy, the values of $A_v,B_v,C_v,E_{uv}$, the gadget spine prefixes, $\Psi_{a,b}$,
  $\Psi'_{a,b}$ and $U_{a,b}$ are pairwise distinct words, and $h$ is injective.
\item The spine tokens $U_{\le j}$ ($j\ge2$) spell $h(W_{(1)}\cdots W_{(j)})$. They have pairwise different lengths, and
  each contains the code of a boundary pair $\gr b\gl{a'}$, which occurs in no single copy; so they differ from all
  per-copy tokens.
\end{itemize}
So no two merges of the schedule produce the same string. In the no case, \cref{lem:merge2slp} holds verbatim for
string-identified merges. This gives the claim for the one-string dataset $\{F\}$.
\end{proof}

\section{Proofs for the algorithm and the barrier}\label{app:gap}

\begin{proof}[Proof of \cref{thm:opeupper}]
Cut $s$ into aligned blocks $b_1\cdots b_r$ of length $m$, the last one possibly shorter, so $r=\lceil n/m\rceil$. For
$\ell=2,\dots,m$, and for each distinct word $u$ of length $\ell$ that is a prefix of some block, do one partial merge of
the pair $(X_{u[1..\ell-1]},u[\ell])$, where $X_v$ is the symbol created for $v$ and $X_v:=v$ if $|v|=1$. The merge acts
on exactly the occurrences at the start of a block with prefix $u$. Invariant before stage $\ell$: every block of length
at least $\ell-1$ reads $X_{b[1..\ell-1]}$ followed by its remaining letters. So the chosen occurrences exist; they lie in
distinct blocks, so they are disjoint and form a legal partial merge, and the invariant holds for $\ell+1$. (A short last
block sits at the end of the string, so it is never followed by a letter of another block.) After stage $m$ every block is
one symbol. The number of merges is the number of distinct block prefixes of length $2$ to $m$, at most
$\sum_{\ell=2}^m\sigma^\ell$. For binary strings this sum is $2^{m+1}-4$, which gives $m=\lfloor\log(k+4)\rfloor-1$; over
$\sigma$ letters it is at most $\sigma^{m+1}/(\sigma-1)\le k$ for $m=\lfloor\log_\sigma k\rfloor-1$.
\end{proof}

\begin{proof}[Proof of \cref{thm:omsupper}]
Let $\mathbf u$ be the current sequence and $\mathcal T$ the set of symbols in it at the start of a phase. Every symbol
is fresh when created, as in the model of Kozma and Voderholzer.
\begin{itemize}
\item \emph{Batch A.} For every pair $(a,b)\in\mathcal T\times\mathcal T$ that occurs adjacently in $\mathbf u$, in any
  order, do the full merge of $(a,b)$. New symbols are not merged within batch A.
\item \emph{Batch B.} For every pair $(a,P)$ with $a\in\mathcal T$ and $P$ a batch-A symbol that occurs adjacently, do the
  full merge of $(a,P)$.
\end{itemize}
\emph{Claim A: after batch A, no two adjacent symbols both lie in $\mathcal T$.} Symbols of $\mathcal T$ are never
re-created, so two adjacent old symbols $a,b$ at the end of batch A were present and adjacent when $(a,b)$ was merged. If
$a\ne b$, occurrences of $ab$ do not overlap and the full merge replaces all of them. If $a=b$, the left-to-right pass
pairs up each run $a^r$ except possibly its last symbol, so two adjacent copies cannot both survive.

\emph{Claim B: after batch B, at most one symbol of $\mathcal T$ remains, and only as the last symbol.} After batch A every
old symbol is followed by a batch-A symbol or is last (Claim A). Pairs of type (old, batch-A) cannot overlap, so the full
merge of $(a,P)$ replaces every such occurrence present; and $P$ cannot have been consumed earlier in batch B, since its
only old left neighbour is $a$, and batch B never pairs $P$ with a symbol on its right.

So every new symbol covers $2$ or $3$ old symbols, except possibly one old symbol at the end, and the length satisfies
$|\mathbf u'|-1\le(|\mathbf u|-1)/2$. After $I$ phases the length is at most $(n-1)/2^I+1$.

\emph{Merge count.} Phase $i$ uses at most $|\mathcal T_i|^2$ merges in batch A and $|\mathcal T_i|^3$ in batch B, and
$|\mathcal T_{i+1}|\le|\mathcal T_i|^2+|\mathcal T_i|^3+1\le3|\mathcal T_i|^3$. With $x_i=\log|\mathcal T_i|$ this gives
$x_{i+1}\le3x_i+\log3$, so $x_i+c\le3^i(x_0+c)$ with $c=(\log3)/2<0.8$. For binary strings $x_0=1$, so
$x_i\le1.8\cdot3^i$ and phase $i$ costs at most $2|\mathcal T_i|^3\le2^{5.4\cdot3^i+1}$. The cost of phases $0,\dots,I-1$
is dominated by its last term up to a factor $2$, so it is at most $2^{1.8\cdot3^I+2}\le k$. A partial last phase only
shortens the sequence. The closed form follows from $2^I\ge\bigl((\log k-2)/1.8\bigr)^{\log_32}/2$. Over $\sigma$ letters
$x_0=\log\sigma$, and $\log\sigma+0.8\le1.8\log\sigma$ for $\sigma\ge2$.
\end{proof}

\begin{proof}[Proof of \cref{thm:barrier}]
We give the proof for $\LENp$ on binary strings. The other cases use the matching bound of \cref{cor:algo} in step (1);
over $\sigma$ letters the hypothesis together with $\rho\le n$ forces $\log\sigma=o(\log n)$, so that bound is
$o(\theta_{\mathrm{no}})$. Step (3) is unchanged, because $\LENf\le\theta$ implies $\LENp\le\theta$, except that the merge
rules are pairs among at most $k+\sigma$ symbols, and $\sigma\le n$ keeps the count $2^{O(n^\varepsilon\log n)}$. Instances $(s,k)$ are encoded with $k$ in
binary. Since $\theta_{\mathrm{no}}\le n$, we have $\theta_{\mathrm{yes}}\le n/\rho=o(\log n)$. Let the reduction $f$ from
SAT run in time at most $N^c$ on inputs of length $N$. Fix $\varepsilon>0$, to be chosen from $c$ only.

(1) \emph{No-instances have small $k$.} By \cref{cor:algo}, every $(s,k)$ with $k\ge n^\varepsilon$ has
$\LENp\le n/(\varepsilon\log n-2)+1<\theta_{\mathrm{no}}(n)$ for $n\ge n_0(\varepsilon)$, because
$\theta_{\mathrm{no}}\ge\rho=\omega(n/\log n)$. So no no-instance of length $n\ge n_0$ has $k\ge n^\varepsilon$.

(2) \emph{A modified reduction.} Let $S_\varepsilon=\{(s,k):k<|s|^\varepsilon,\ \LENp(s,k)\le\theta_{\mathrm{yes}}(|s|)\}$.
If $S_\varepsilon$ is empty, then for $n\ge n_0$ a SAT image is a yes-instance exactly when $k\ge|s|^\varepsilon$, so
SAT is in P and the conclusion is trivial. Otherwise fix a member $y_0$ of $S_\varepsilon$ and a non-member $y_1$. Let
$f'(\varphi)$ be: if $f(\varphi)=(s,k)$ with $|s|<n_0$, decide $\varphi$ by brute force on $f(\varphi)$ (finitely many
strings) and output $y_0$ or $y_1$; else, if $k\ge|s|^\varepsilon$, output $y_0$; else output $f(\varphi)$. Then $f'$ is a
Karp reduction from SAT to $S_\varepsilon$: a satisfiable $\varphi$ maps to $y_0$ or to a yes-instance with
$k<n^\varepsilon$, and an unsatisfiable $\varphi$ maps, by (1), to $f(\varphi)$ itself, whose optimum is at least
$\theta_{\mathrm{no}}>\theta_{\mathrm{yes}}$. Its running time is $O(N^c)$, independent of $\varepsilon$ up to the
constant-size table for $|s|<n_0$.

(3) \emph{Density.} Partial merges preserve the spelled string, so $s$ is determined by its $k$ merge rules (each a pair
among at most $k+2$ symbols) and the final sequence of at most $\theta_{\mathrm{yes}}$ symbols. Hence
$|S_\varepsilon\cap\{|s|=n\}|\le n^\varepsilon(k+2)^{2k}(k+2)^{\theta_{\mathrm{yes}}+1}\le2^{C n^\varepsilon\log n}$, and
the number of members of length at most $N$ is at most $2^{C'N^\varepsilon\log N}$.

(4) \emph{Density theorem.} Complementing, $f'$ is a many-one reduction from coSAT to the complement of $S_\varepsilon$.
By Buhrman and Hitchcock \cite[Lemma 3.5]{buhrman2008dense}, the complement of a set of density $d$ disjunctively
reduces to a set $T$ of density at most $N\,d(N)+N$. The composed disjunctive reduction from coSAT to $T$ runs in time
$q(N)\le N^{c'}$, with $c'$ depending only on $c$. We use Lemma~3.2 of \cite{buhrman2008dense}, which states: if a set $A$ has an AND-function of order
$s$ (in the sense of \cite{buhrman2008dense}), $\alpha<1/s$, and $A$ NP-reduces to a family of sets $B=(B_n)$ with
$\|B_n\|\le2^{n^\alpha}$ for all sufficiently large $n$, then $A\in\mathrm{NP/poly}$. An AND-function for coSAT is an
OR-function for SAT; let $s_0$ be its order (order $1$ for the standard one). Choose $\varepsilon<1/(2s_0c')$. Then the part of $T$ queried on
inputs of length $N$ has at most $q\cdot2^{C'q^\varepsilon\log q}+q\le2^{N^{\alpha}}$ elements for large $N$, with
$\alpha<1/s_0$. Lemma~3.2 of \cite{buhrman2008dense}, applied with $A=$ coSAT and $B_N$ this part of $T$, then gives
coSAT $\in\mathrm{NP/poly}$, that is,
$\mathrm{coNP}\subseteq\mathrm{NP/poly}$. The order of quantifiers matters: $c$ and $c'$ are fixed by $f$ before
$\varepsilon$ is chosen. Lemma~3.2 places no computational requirement on $T$.
\end{proof}

\section{Proofs for the unary results}\label{app:unary}

\subsection{OPE-seg tokenisation}
We use the construction and notation of \cref{sec:unary}.

\paragraph{Facts about the numbers.}

\begin{lemma}[Sidon property]\label{lem:sidon}
For $1\le i\le j\le|V|$ the sums $\xi_i+\xi_j$ are pairwise distinct; hence so are the sums $\theta_i+\theta_j$.
\end{lemma}

\begin{proof}
Let $\Xi=\xi_i+\xi_j=2\varpi(i+j)+(i^2\bmod\varpi)+(j^2\bmod\varpi)$. The last two terms sum to at most
$2\varpi-2<2\varpi$, so $i+j=\lfloor\Xi/2\varpi\rfloor$ is determined by $\Xi$, and so is
$i^2+j^2\equiv\Xi-2\varpi(i+j)\pmod\varpi$. Then $(i-j)^2=2(i^2+j^2)-(i+j)^2$ is determined modulo $\varpi$, so
$i-j\equiv\pm\Delta_0\pmod\varpi$ for a determined $\Delta_0$. With $i+j$ known and $\varpi$ odd, $\{i,j\}$ is
determined modulo $\varpi$, and since $1\le i,j\le|V|<\varpi$ it is determined. Multiplying by $3$ keeps sums distinct.
\end{proof}

\begin{lemma}[ranges]\label{lem:ranges}
$0<\theta_v<U/4$ for all $v$. Hence $\Pi\subset[U,\frac54U)$ and
$\mathcal A\subset[9U,\frac{37}4U)$; $\mathcal S\cap[1,U)$ consists of powers of two only;
$\mathcal S\cap[\frac54U,9U)=\{2U,4U,8U\}$; $\mathcal S\cap[9U,M]=\mathcal A\cup\{16U,32U,64U,128U\}$; $\max\mathcal
S=M$, and $M$ is the only element of $\mathcal S$ in $(M/2,M]$, since $\frac{37}4U<64U=M/2$.
\end{lemma}

\begin{proof}
$\theta_v<U/4$ is the choice of $\beta$. Every $\phi_{v,i}$ contains the bit $U$ and some lower bits summing to less
than $U/4$, and $\alpha_v=9U+\theta_v$; the remaining claims list the powers of two in each range.
\end{proof}

\begin{lemma}[scaffold chain]\label{lem:chain}
$\mathcal S$ can be built with exactly $|\mathcal S|-1$ merges.
\end{lemma}

\begin{proof} $2^i=2^{i-1}+2^{i-1}$
builds $\mathcal P$; $\phi_{v,1}=U+2^{b_1}$ and $\phi_{v,i}=\phi_{v,i-1}+2^{b_i}$ build $\Pi$; and $\alpha_v=8U+\phi_{v,r}$
builds $\mathcal A$, with $8U\in\mathcal P$. Each element of the set $\mathcal S\setminus\{1\}$ costs one merge
(prefix sums of different vertices may coincide; they are built once).
\end{proof}

Let $Z_{\mathrm{cov}}=\{\alpha_v+M\}$ and $Z_{\mathrm{edge}}=\{\ell_e\}$. Note $\alpha_v+M\in[M+9U,M+\frac{37}4U)$ and
$\ell_e=M+18U+\theta_u+\theta_v\in[M+18U,M+\frac{37}2U)$.

\begin{lemma}\label{lem:Lfacts}
\begin{enumerate}\renewcommand{\labelenumi}{(L\arabic{enumi})}
\item $Z_{\mathrm{cov}}\cap\mathcal S=\emptyset$.
\item For every edge $e$: $\ell_e\notin\mathcal S$, and $\ell_e$ is not $x+y$ with $x,y\in\mathcal S$.
\item For an edge $e=\{u,v\}$ and a vertex $w'\notin e$: $\ell_e-(\alpha_{w'}+M)\notin\mathcal S$. (For $w'=u$ it equals
  $\alpha_v\in\mathcal S$.)
\item For distinct edges $e,e'$: $\ell_e-\ell_{e'}\notin\mathcal S\cup\{0\}$.
\item $Z_{\mathrm{cov}}\cap Z_{\mathrm{edge}}=\emptyset$; $v\mapsto\alpha_v+M$ and $e\mapsto\ell_e$ are injective; and
  a sum of two elements of $Z_{\mathrm{cov}}\cup Z_{\mathrm{edge}}$ exceeds every $\ell_e$.
\end{enumerate}
\end{lemma}

\begin{proof}
(L1) Every element of $Z_{\mathrm{cov}}$ exceeds $M=\max\mathcal S$.
(L2) $\ell_e>M$. If $\ell_e=x+y$ with $x\ge y$ in $\mathcal S$, then $x\ge\ell_e/2>M/2$, so $x=M$ by \cref{lem:ranges}, and
$y=18U+\theta_u+\theta_v\in[18U,\frac{37}2U)$; by \cref{lem:ranges}, $\mathcal S$ has no element there.
(L3) $\ell_e-\alpha_{w'}-M=9U+(\theta_u+\theta_v-\theta_{w'})$ with $\theta_u+\theta_v-\theta_{w'}\in(-U/4,U/2)$, so
the value lies in $(\frac{35}4U,\frac{19}2U)$. By \cref{lem:ranges}, $\mathcal S$ meets this interval only in $\mathcal A$. If the
value were $\alpha_x=9U+\theta_x$, then $\theta_u+\theta_v=\theta_{w'}+\theta_x$, so $\{u,v\}=\{w',x\}$ by
\cref{lem:sidon}, and $w'\in e$, a contradiction.
(L4) $\ell_e-\ell_{e'}=\theta_u+\theta_v-\theta_{u'}-\theta_{v'}$ is nonzero by \cref{lem:sidon}, has absolute value less
than $U/2$, and is divisible by $3$. A positive element of $\mathcal S$ below $U/2$ is a power of two, which is not
divisible by $3$.
(L5) $\alpha_v+M<M+\frac{37}4U<M+18U\le\ell_e$; injectivity follows from \cref{lem:sidon} (and distinctness of the
$\theta_v$); any sum of two such elements is at least $2M+18U>\ell_e$.
\end{proof}

\paragraph{($\Rightarrow$)} Let $C$ be a vertex cover with $|C|\le\kappa$; enlarge it arbitrarily to $|C|=\kappa$. Build
$\mathcal S$ ($|\mathcal S|-1$ merges, \cref{lem:chain}), then merge $(\alpha_v,M)$ for $v\in C$ ($\kappa$ merges, both parts in
$\mathcal S$). Every $D_1$ string is a token (cost $1$). A cover string costs $1$ if $v\in C$ and at most $2$ otherwise.
An edge string $\{u,v\}$ with, say, $u\in C$ costs at most $2$, as $(\alpha_u+M)+\alpha_v$. The total is at most
$\omega(|\mathcal S|-1)+(2|V|-\kappa)+2|E|=\delta$.

\paragraph{($\Leftarrow$)} Let a merge sequence of length $k$ have vocabulary $\mathcal V$ and total cost at most
$\delta$.

\emph{Step 1: $\mathcal S\subseteq\mathcal V$.} Otherwise some $z\in\mathcal S\setminus\{1\}$ is not a token, its
$\omega$ copies cost at least $2$ each, and every string costs at least $1$, so the total is at least
$\omega(|\mathcal S|-1)+\omega+|V|+|E|>\delta$, because $\omega=|V|+|E|+1$.

\emph{Step 2.} Each merge adds at most one token, so $|\mathcal V|\le k+1$, and $\mathcal Z:=\mathcal V\setminus\mathcal S$
has $|\mathcal Z|\le\kappa$.

\emph{Step 3 (counting).} Every $D_1$ string costs exactly $1$. A cover string costs at most $2$ (as $\alpha_v,M\in\mathcal
S$), and costs $1$ if and only if $\alpha_v+M\in\mathcal Z$ (L1). An edge string costs at most $3$, and costs $1$ if and only
if $\ell_e\in\mathcal Z$ (L2). Let $S_0=\{v:\alpha_v+M\in\mathcal Z\}$, $E_1=\{e:\ell_e\in\mathcal Z\}$ and
$c_2=|\{e:\cost_{\mathcal V}(\ell_e)\le2\}|$. Then $c_2\ge|E_1|$ and the total cost is
$\omega(|\mathcal S|-1)+(2|V|-|S_0|)+(3|E|-c_2-|E_1|)$. So the cost bound gives
\begin{equation}\label{eq:u1}
|S_0|+c_2+|E_1|\ \ge\ \kappa+|E| .
\end{equation}
By (L5) the tokens $\alpha_v+M$ ($v\in S_0$) and $\ell_e$ ($e\in E_1$) are distinct elements of $\mathcal Z$, so
\begin{equation}\label{eq:u2}
|S_0|+|E_1|\ \le\ |\mathcal Z|\ \le\ \kappa .
\end{equation}
From \eqref{eq:u1} and \eqref{eq:u2}, $c_2\ge|E|$; hence $c_2=|E|$, equality holds in \eqref{eq:u2}, and $\mathcal Z$
consists exactly of the cover tokens of $S_0$ and the edge tokens of $E_1$.

\emph{Step 4 (every edge is covered by $S_0$).} Fix $e=\{u,v\}$. Since $\cost_{\mathcal V}(\ell_e)\le2$, either
$\ell_e=x+y$ with $x,y\in\mathcal V$, or $\ell_e\in\mathcal V$; in the second case $\ell_e\in\mathcal Z$ was created by
a merge of two earlier tokens summing to $\ell_e$. So in all cases $\ell_e=x+y$ with $x,y\in\mathcal V$. By (L2) not both
are in $\mathcal S$; by (L5) and Step 3 not both are in $\mathcal Z$. So, say, $x\in\mathcal Z$ and $y\in\mathcal S$. If
$x=\ell_f$ with $f\in E_1$, then $f\ne e$ (as $y>0$) and $y=\ell_e-\ell_f\in\mathcal S$ contradicts (L4). Hence
$x=\alpha_{w'}+M$ with $w'\in S_0$, and $y=\ell_e-(\alpha_{w'}+M)\in\mathcal S$ forces $w'\in e$ by (L3).

So $S_0$ is a vertex cover of size at most $|\mathcal Z|\le\kappa$. Since all lengths are $O(|V|^2)$ and the dataset is
written in unary, this is strong NP-hardness. Together with membership this proves \cref{thm:unary}. \hfill$\square$

\subsection{The two readings of the OPE-seg model}

\begin{remark}[two readings of the model]\label{rem:readings}
Kastreva et al.\ define the extracted vocabulary as $\{\str a\}\cup\{s_1\circ s_2:(s_1,s_2)\in\mathbf m\}$, with merges
drawn from pairs of strings, and their definition does not say whether both parts of a merge must already be tokens.
\begin{enumerate}\renewcommand{\labelenumi}{(\alph{enumi})}
\item \emph{Literal reading} (any pair of strings). Then the vocabulary is $\{\str a\}$ plus any $k$ strings of length at
  least $2$, so unary OPE-seg coincides with unary direct tokenisation and is strongly NP-complete by their Theorem~5;
  \cref{thm:unary} is then a trivial corollary. Under this reading their Theorem~6 proof would not go through: for a
  single target $\ell\ge3$, one merge would make $\ell$ a token, but no addition chain of length $1$ reaches $\ell$.
\item \emph{Reachable reading} (both parts already built), used above. Their proof of Theorem~6 mentions non-reachable
  merges only in a footnote (their Appendix~N, footnote~9), which says that subwords produced by non-reachable merges
  ``should be pruned'' from the addition chain read off the vocabulary. Pruning is harmless only if such subwords cannot
  be tokens that the encoding uses, so their proof goes through under this reading, and under it \cref{thm:unary} is
  new. It also matches Kozma and Voderholzer: their merges act on pairs of symbols, and a merge of symbols that do not
  occur has no effect. In the unary case, applying the merges as partial merges to each string gives exactly the costs
  of optimal segmentation with the vocabulary of buildable tokens.
\end{enumerate}
So unary OPE-seg tokenisation is strongly NP-complete under either reading; it is new under the reachable one.
\end{remark}

\subsection{Standard bottom-up tokenisation}
\begin{proof}[Proof of \cref{lem:valuation}]
By induction on merges. Initially the string is $1^\ell$: $r=\ell$, no tail, $j=0$. The adjacent pairs of
$(2^j)^ru_1\cdots u_c$ are $(2^j,2^j)$ (only inside the head run, since no $u_i$ equals $2^j$), $(2^j,u_1)$ (if $r\ge1$
and $c\ge1$), and $(u_i,u_{i+1})$. Consider a merge of a pair of tokens.
\begin{itemize}
\item \emph{Doubling $(2^j,2^j)$.} It acts exactly on the head run: $(2^j)^r$ becomes
  $(2^{j+1})^{\lfloor r/2\rfloor}(2^j)^{r\bmod2}$, and the tail is untouched. The new head is $2^{j+1}$, and the new tail
  $(2^j)^{r\bmod2}u_1\cdots u_c$ has strictly decreasing valuations, all below $j+1$. (If $r\le1$ the merge does not act
  on this string, and the same description holds.)
\item \emph{The pair $(2^j,u_1)$.} It is non-doubling since $u_1\ne2^j$, and it occurs once, at the head--tail boundary.
  The result is $(2^j)^{r-1}(2^j+u_1)u_2\cdots$, and $\mathrm{ord}_2(2^j+u_1)=\mathrm{ord}_2(u_1)$.
\item \emph{A pair $(u_i,u_{i+1})$.} The $u_i$ are distinct, so it occurs once. The new token has valuation
  $\mathrm{ord}_2(u_{i+1})$, so valuations stay strictly decreasing and below $j$; it is not divisible by $2^j$, so it
  is not absorbed into the head.
\item Any other merge does not occur in the string.\qedhere
\end{itemize}
\end{proof}

\begin{proof}[Proof of \cref{lem:potential}]
Initially equality holds. A doubling maps $r\mapsto\lfloor r/2\rfloor$, $c\mapsto c+(r\bmod2)$, $j\mapsto j+1$, and
for $j'\ge j+1$ the left side is unchanged by the identity for $\mathrm{pc}$. A boundary merge maps $r\mapsto r-1$ and
$\iota\mapsto\iota+1$; the left side drops by at most $1$, because $\mathrm{pc}_i(r)-\mathrm{pc}_i(r-1)\le1$. A tail
merge lowers $c$ by exactly $1$ and raises $\iota$ by $1$. A merge that does not act changes nothing.
\end{proof}

\begin{proof}[Proof of \cref{lem:single}]
\emph{Lower bound.} Let $j\le k$ be the number of doublings. At most $k-j$ non-doubling merges act on $\ell$, so
\cref{lem:potential} with $j'=j$ gives final length at least $\mathrm{pc}_j(\ell)-(k-j)$. Since
$\mathrm{pc}_j(\ell)-\mathrm{pc}_{j+1}(\ell)=\lfloor\ell/2^{j+1}\rfloor$, the function $j\mapsto\mathrm{pc}_j(\ell)+j$ is
non-increasing while $2^{j+1}\le\ell$, and $\mathrm{pc}_j(\ell)$ is constant once $2^{j+1}>\ell$. Hence
$\mathrm{pc}_j(\ell)+j-k\ge\mathrm{pc}_J(\ell)+J-k$ for every $j\le k$. Also the length is at least $1$.

\emph{Upper bound.} $J$ doublings turn $1^\ell$ into its binary expansion: $\lfloor\ell/2^J\rfloor$ copies of $2^J$
followed by the powers of two of $\ell\bmod2^J$ in decreasing order (the doubling step of \cref{lem:valuation}), of
length $\mathrm{pc}_J(\ell)$. Each of the remaining $k-J$ merges, chosen as any pair present while the length is at
least $2$, shortens the string by at least $1$. Once the length is $1$, pad with merges that act nowhere.
\end{proof}

\paragraph{Proof of \cref{thm:unaryfull}.} We use the construction and notation of \cref{sec:unaryfull}.

\paragraph{($\Rightarrow$) A biclique gives total length $\delta$.} Let $S_1\times S_2\subseteq E$ with
$|S_1|=|S_2|=\kappa$. Use the following $|\mathcal S|-1+2\kappa=k$ merges.
\begin{enumerate}
\item The doublings $(2^i,2^i)$ for $i=0,\dots,h+2d_1-1$.
\item For every node of the side-$1$ trie, in order of size $2,3,\dots$: the merge (value of the prefix of size $i-1$,
  $2^b$), where $b$ is the $i$-th bit of the code; then the same for the side-$2$ trie. One merge per element of
  $\mathcal T_1\cup\mathcal T_2$.
\item $(c^{(1)}_x,Q_1)$ for $x\in S_1$, then $(Q_2,c^{(2)}_y)$ for $y\in S_2$.
\item $(Q_1,Q_2)$.
\end{enumerate}
After step 1 every string, being less than $2^{h+2d_1+1}$, is its binary expansion in decreasing order
(\cref{lem:valuation}). In step 2 we keep the invariant that after all trie merges of size at most $i$, the high part of
every string (its bits in the high block and the high leading bit, which is either a single bit or a prefix of a code)
reads as the token of its first $\min(i,\text{size})$ bits followed by its remaining bits one by one. A trie merge
(value of $\varphi$, $2^b$) with $|\varphi|\ge1$ occurs exactly where a high part starts with $\varphi$ followed by $b$: only the first
token of a high part can contain the high leading bit; prefix tokens have at least two bits and so equal no other token;
and the element after a complete high part is a bit outside the high block. The same holds for the low block with
leading bit $2d_2$, and the two tries never touch each other's blocks. After step 2, every $D_1$ string in
$\mathcal P\cup\mathcal T_1\cup\mathcal T_2$ is a single token; the string $Q_{12}$ reads $Q_1,Q_2$; the vertex strings read
$c^{(1)}_x,Q_1$ and $Q_2,c^{(2)}_y$; and the edge strings read $c^{(1)}_x,Q_1,Q_2,c^{(2)}_y$.

In step 3, $(c^{(1)}_x,Q_1)$ turns the vertex string of $x$ into one token and the edge strings at $x$ into
$(c^{(1)}_x{+}Q_1)$, $Q_2$, $c^{(2)}_y$; then $(Q_2,c^{(2)}_y)$ acts on the vertex string of $y$ and on every edge string
ending in $Q_2$, $c^{(2)}_y$ (step 3 never consumes the $Q_2$ of an edge string through a merge whose first part is
$c^{(1)}_x$). Step 4 acts on the string $Q_{12}$ and on the edge strings in which neither $Q_1$ nor $Q_2$ was consumed.
Final lengths: $D_1$ strings $1$; vertex strings $1$ if chosen and $2$ otherwise; an edge string $(x,y)$ has $2$ tokens
if $x\in S_1$ and $y\in S_2$, and $3$ otherwise (\cref{fig:biclique}). The total is
$\omega_0(|\mathcal S|-1)+\omega_1(2(|V_1|+|V_2|)-2\kappa)+3|E|-e(S_1,S_2)=\delta$, where $e(S_1,S_2)=\kappa^2$ is the
number of edges between $S_1$ and $S_2$.

\paragraph{($\Leftarrow$) Total length at most $\delta$ gives a biclique.} Let $|\mathbf m|\le k$ have total length at
most $\delta$, put $\mathcal V=\mathcal V(\mathbf m)$, and use \eqref{eq:LB} throughout.

\emph{Step 1: $\mathcal S\subseteq\mathcal V$.} Otherwise some $z\in\mathcal S\setminus\{1\}$ is missing, its
$\omega_0$ copies cost at least $2$ each, every other string costs at least $1$, and the total is at least
$\omega_0(|\mathcal S|-1)+\omega_0+\omega_1(|V_1|+|V_2|)+|E|>\delta$ by the choice of $\omega_0$.

\emph{Step 2.} $|\mathcal V|\le k+1$, so at most $2\kappa$ elements of $\mathcal V$ lie outside $\mathcal S$. Let
$V_1^\ast=\{x:c^{(1)}_x+Q_1\in\mathcal V\}$ and $V_2^\ast=\{y:Q_2+c^{(2)}_y\in\mathcal V\}$. These tokens are not in $\mathcal S$
(bit-set view) and are pairwise distinct, so $|V_1^\ast|+|V_2^\ast|\le2\kappa$.

\emph{Step 3: counting.} The $D_1$ strings cost at least $\omega_0(|\mathcal S|-1)$. A vertex string costs at least $2$
unless it is in $\mathcal V$, so $D_2$ costs at least $\omega_1(2(|V_1|+|V_2|)-|V_1^\ast|-|V_2^\ast|)$. Write the cost of an edge
string as $3-\mathrm{sav}_e$ with $\mathrm{sav}_e\le2$. The bound $\delta$ gives
$\omega_1(|V_1^\ast|+|V_2^\ast|)+\sum_e\mathrm{sav}_e\ge2\kappa\omega_1+\kappa^2$. If $|V_1^\ast|+|V_2^\ast|\le2\kappa-1$, the left side is
at most $\omega_1(2\kappa-1)+2|E|<2\kappa\omega_1$. Hence $|V_1^\ast|+|V_2^\ast|=2\kappa$, so
$\mathcal V=\mathcal S\cup\{c^{(1)}_x+Q_1:x\in V_1^\ast\}\cup\{Q_2+c^{(2)}_y:y\in V_2^\ast\}$, and $\sum_e\mathrm{sav}_e\ge\kappa^2$.

\begin{lemma}\label{lem:edgecost}
For this $\mathcal V$ and an edge $e=(x,y)$: $\cost_{\mathcal V}(\ell_e)=2$ if $x\in V_1^\ast$ and $y\in V_2^\ast$, and
$\cost_{\mathcal V}(\ell_e)\ge3$ otherwise.
\end{lemma}

\begin{proof}
$\ell_e$ has bits in the high block, at $\mu_1$, at $\mu_2$ and in the low block; no element of $\mathcal V$ does, so
$\ell_e\notin\mathcal V$. Suppose $\ell_e=\tau+\tau'$ with $\tau,\tau'\in\mathcal V$, and write $\bar z=z\bmod2^{\mu_2}$. For
$z\in\mathcal V$: $\bar z=z$ if $z<2^{\mu_2}$ (powers $2^i$ with $i\le2d_2+1$, and $\mathcal T_2$);
$\overline{Q_2+c^{(2)}_{y'}}=c^{(2)}_{y'}$; and $\bar z=0$ for all other $z$ (larger powers, $\mathcal T_1$, $Q_{12}$, and
$c^{(1)}_{x'}+Q_1$). Every residue is at most $2^{2d_2+1}=2^{\mu_2}/2$, so $\bar\tau+\bar\tau'\le2^{\mu_2}$, with equality only if
both equal $2^{2d_2+1}$, in which case the sum is $0$ modulo $2^{\mu_2}$. Since $\ell_e\equiv c^{(2)}_y$ and
$0<c^{(2)}_y<2^{\mu_2}$, we get $\bar\tau+\bar\tau'=c^{(2)}_y$ exactly. If $\tau,\tau'<2^{\mu_2}$ then $\tau+\tau'<2^{\mu_2+1}\le2^{h+2d_1}\le\ell_e$;
so say $\tau\ge2^{\mu_2}$.
\begin{itemize}
\item If $\bar\tau=0$, then $\bar\tau'=c^{(2)}_y$, so $\tau'=c^{(2)}_y$ (if $\tau'<2^{\mu_2}$) or $\tau'=Q_2+c^{(2)}_y$ (the only element at
  least $2^{\mu_2}$ with that residue). If $\tau'=c^{(2)}_y$ then $\tau$ has bit set $\mathrm{code}_1(x)\cup\{\mu_1,\mu_2\}$, which is
  not in $\mathcal V$. If $\tau'=Q_2+c^{(2)}_y$ then $\tau=c^{(1)}_x+Q_1$; and $\tau\in\mathcal V$ iff $x\in V_1^\ast$, $\tau'\in\mathcal
  V$ iff $y\in V_2^\ast$.
\item If $\tau=Q_2+c^{(2)}_{y'}$ with $y'\in V_2^\ast$, then $\bar\tau'=c^{(2)}_y-c^{(2)}_{y'}$. If $\tau'<2^{\mu_2}$, then
  $\tau+\tau'=Q_2+c^{(2)}_y<\ell_e$. If $\tau'\ge2^{\mu_2}$, then $\bar\tau'\in\{0\}\cup\{c^{(2)}_{y''}\}$; the second option is excluded
  because $c^{(2)}_{y'}+c^{(2)}_{y''}\ge2^{2d_2+1}>c^{(2)}_y$ (every low code contains bit $2d_2$). So $\bar\tau'=0$, $y'=y$,
  and $\tau'=c^{(1)}_x+Q_1$, which is in $\mathcal V$ only if $x\in V_1^\ast$.\qedhere
\end{itemize}
\end{proof}

By \cref{lem:edgecost}, $\mathrm{sav}_e\le[x\in V_1^\ast][y\in V_2^\ast]$, so the number of edges between $V_1^\ast$ and
$V_2^\ast$ is at least $\sum_e\mathrm{sav}_e\ge\kappa^2$, while $|V_1^\ast|+|V_2^\ast|=2\kappa$. Since that number is at most
$|V_1^\ast|\,|V_2^\ast|\le\kappa^2$, equality holds throughout: $|V_1^\ast|=|V_2^\ast|=\kappa$ and $V_1^\ast\times V_2^\ast\subseteq E$.
This proves \cref{thm:unaryfull}. \hfill$\square$

\section{Reproducibility}\label[appendix]{app:repro}
\sloppy

The scripts and result files for all checks below are in the supplementary archive deposited with this preprint on Zenodo
(\href{https://doi.org/10.5281/zenodo.24023730}{doi:10.5281/zenodo.24023730}); paths refer to that archive, whose README also lists the full per-run tables. All proofs are complete
without computation; the checks guard against errors in the proofs and constructions. Every number not following from a displayed formula comes from a file named here.
Checks use exact integer or rational arithmetic unless marked otherwise.

\paragraph{Unbounded-alphabet gadget (\cref{sec:warmup}).} \nolinkurl{gadget/verify.py} $\to$
\nolinkurl{gadget/verify_results.json} checks \cref{lem:gadget} with an exact smallest-SLP solver on $50$ graphs ($11$
named and $39$ random, at most $8$ vertices), and replays \cref{lem:sched} (final length $1$ after exactly $R(|S|)$ full
merges) on those graphs and on $22$ cubic graphs with up to $30$ vertices. \nolinkurl{binary/cor4_check.py} $\to$
\nolinkurl{binary/cor4_check.json} checks \cref{cor:gsH} exactly on whole gadget strings with a parse-tree integer program;
the pass condition $g(s_H)=R(\vc(H))$ was fixed in advance and holds on $9$ graphs: $K_2$ ($|s_H|=18$, $g=15$), $P_3$
($30$, $24$), $K_3$ ($36$, $28$), $K_{1,3}$ ($42$, $33$), $P_4$ ($42$, $34$), $C_4$ ($48$, $37$), the paw ($48$, $37$),
the diamond ($54$, $40$) and $K_4$ ($60$, $44$); larger graphs did not finish. A script checks \cref{lem:realise} on $40$
random binary and ternary strings of length at most $9$ (\nolinkurl{gadget/check_partial.py} $\to$
\nolinkurl{gadget/check_partial_results.json}).

\paragraph{Binary construction (\cref{sec:binary}).} \nolinkurl{binary/oms_schedule.py} $\to$ \nolinkurl{binary/oms_schedule.json}
builds $F$ bit by bit, simulates full merges on it, and checks six conditions fixed before the runs: final length $1$; no
identity merge; the Stage-A image is the codeword-token image of $W$; exactly $\Ccode+t(R(|S|)+3)-1$ merges;
$\Ccode\le3+(2d+1)q_W$; all created tokens spell distinct strings. All $37$ runs pass ($14$ graphs with up to $14$
vertices, up to $t=p^2=169$ copies, up to $224{,}770$ bits). A second implementation sharing no code,
\nolinkurl{independent_check/oms_sim.py} $\to$ \nolinkurl{independent_check/oms_sim.json}, randomises letter numbering, inner
coordinates, word order and trie-merge order, and also checks that trie merges fire only at codeword starts and Stage-B
merges stay in their blocks; all $160$ configurations pass (primes $11$ to $47$, up to $289$ copies, up to $428{,}298$
bits), and one control run with a non-minimal cover produces an identity merge, as expected. A mutation test
(\nolinkurl{independent_check/mutation_test.py} $\to$ \nolinkurl{independent_check/mutation_test.json}) shows the checks are not vacuous:
identical copies without the affine relabelling, or the code $\eta(0)=\str{00}$, make it fail.
\nolinkurl{binary/lb_checks.py} $\to$ \nolinkurl{binary/lb_checks.json} checks two-separation (\cref{lem:affine}) for $K_2$
($t=121$), $P_3$ ($169$), $C_4$ ($289$), the Petersen graph ($300$) and others, and synchronisation (\cref{lem:sync})
exhaustively for all windows of length $2$ to $3L$ on $K_2$ with $t=7$, $K_3$ with $t=4$ and $K_4$ with $t=3$
($152{,}610$, $165{,}894$ and $203{,}532$ windows). $g(F)$ itself and \cref{lem:directsum,lem:decode} as general
statements are not checked.

\paragraph{Algorithm and barrier (\cref{sec:gap}).} \nolinkurl{gap/sim_gap.py} $\to$ \nolinkurl{gap/sim_gap.json} is a
simulator written separately from the proofs. It applies the construction of \cref{thm:opeupper} with real partial
merges, asserting that every replaced position is an occurrence and that replaced positions do not overlap: $1{,}088$
runs over $\sigma\in\{2,3,4\}$, nine families of strings and lengths up to $30{,}001$, all with exactly $\lceil n/m\rceil$
symbols left and at most $\sum_{\ell=2}^m\sigma^\ell$ merges. It runs the phases of \cref{thm:omsupper} with full merges and
checks Claims~A and~B, the halving of the length, the growth of the symbol set and the cumulative merge budget in every
phase: $378$ runs (six order runs per string: five named orders, the random one twice), all binary strings of length at most $16$ under two orders ($262{,}140$ runs), and
$180$ runs with truncated budgets, without violation. An exact brute force over $1{,}142$ binary strings confirms
\cref{lem:lenlower}. The proofs do not rely on these runs, and \cref{thm:barrier} has no computational component.

\paragraph{Unary OPE-seg (\cref{thm:unary}).} \nolinkurl{unary_ope/reduction_check.py} $\to$
\nolinkurl{unary_ope/reduction_check.json} checks (L1)--(L5) and the Sidon property on six graphs and, for
$\kappa\in\{\vc-1,\vc\}$, minimises the cost of $D_2+D_3$ over all sets of extra buildable tokens, given the scaffold. An
independent program written from the proof text, \nolinkurl{independent_check/unary_ope_check.py} $\to$
\nolinkurl{independent_check/unary_ope_check_log.jsonl}, repeats this on twelve graphs with a full change-making re-check
of each optimum. In every case the minimum is at most $\delta$ exactly when $\vc\le\kappa$, with equality at $\kappa=\vc$
(for $\vc=3$ only the no side was searched).

\paragraph{Unary full merges (\cref{lem:single,thm:unaryfull}).} \nolinkurl{unary_full/single_string.py} $\to$
\nolinkurl{unary_full/single_string_600_12.json} compares \cref{lem:single} with brute force for $\ell\le600$, $k\le12$
($7{,}800$ cases, no mismatch). \nolinkurl{unary_full/single_invariants.py} $\to$
\nolinkurl{unary_full/single_invariants_400_13.json} enumerates all reachable single-string states for $\ell\le400$, $k\le13$
($738{,}666$ states) and checks \cref{lem:valuation,lem:potential}. \nolinkurl{unary_full/biclique_reduction.py} $\to$
\nolinkurl{unary_full/biclique_reduction.json} builds the reduction for seven bipartite graphs and every $\kappa$ ($19$
cases): it simulates the ($\Rightarrow$) sequence for every $(S_1,S_2)$ and confirms the total
$\delta+\kappa^2-e(S_1,S_2)$, checks \cref{lem:edgecost} for every $V_1^\ast,V_2^\ast$, and confirms that the best intended
sequence reaches $\delta$ exactly when a balanced biclique exists; for $\kappa=1$ on $K_{2,2}$, $P_4$, $2K_2$ a search over
all sets of at most two extra integers finds exactly the intended optimum. A second implementation,
\nolinkurl{independent_check/unary_full_check.py}, re-checks \cref{lem:single} for $\ell\le256$, $k\le13$ ($3{,}584$ cases,
\nolinkurl{independent_check/unary_full/thmB3.json}) and \cref{lem:edgecost}, and runs heuristic searches (greedy, beam,
simulated annealing, one- and two-step edits of the intended sequences) on five no-instances, none of which reaches
$\delta$ (\nolinkurl{independent_check/unary_full/s_*.json}, which also holds two yes-instance controls; evidence only). The example in \cref{sec:open} comes from
\nolinkurl{unary_full/bpe_vs_opt.py} (exact search \nolinkurl{unary_full/unary_oms_exact.py}) $\to$
\nolinkurl{unary_full/bpe_vs_opt.json}.
}

\end{document}